\documentclass[12pt]{article}
\usepackage{color, amsmath, amsthm}

\usepackage{graphicx}
\usepackage{amsfonts}
\usepackage{amssymb,amsmath,amscd}
\usepackage{fancyhdr}
\usepackage[top=2.75cm, bottom=2.75cm, left=2.75cm, right=2.75cm]{geometry}

\usepackage{hyperref}

\hypersetup{
    pdftitle={Cohomological localization of Chern-Simons theory},    % title
    pdfauthor={ Johan Kallen},     % author
    pdfnewwindow=true,      % links in new window
    colorlinks=true,       % false: boxed links; true: colored links
    linkcolor=black,          % color of internal links
    citecolor= black,        % color of links to bibliography
    filecolor= black,      % color of file links
    urlcolor= black           % color of external links
}

\def\+{{+\!\!\!+}}

\def\pmb#1{\setbox0=\hbox{#1}% 
\kern.0em\copy0\kern-\wd0 
\kern-.04em\copy0\kern-\wd0 
\kern.08em\copy0\kern-\wd0 
\kern-.04em\raise.0433em\box0 }         %poor man's bold macro (TexBook) 
\newcommand{\nc}{\newcommand} 
\nc{\beq}{\begin{equation}} 
\nc{\eeq}[1]{\label{#1}\end{equation}} 
\nc{\ber}{\begin{eqnarray}} 
\nc{\eer}[1]{\label{#1}\end{eqnarray}} 
\nc{\pek}[1]{\cite{#1}} 
\nc{\enr}[1]{(\ref{#1})} 
\nc{\kal}[1]{{\cal{#1}}} 
\nc{\dott}{\;\cdot\;} 
\nc{\coker}{\mathrm{coker}}
\nc{\ie}{{\it i.e.}}
\nc{\eg}{{\it e.g.}}
\numberwithin{equation}{section}

\def\0 {\nonumber}

\newcommand{\be}{\begin{equation}} 
\newcommand{\ee}{\end{equation}} 
\newcommand{\bea}{\begin{eqnarray}} 
\newcommand{\eea}{\end{eqnarray}}

\begin{document}

\title{Cohomological localization of Chern-Simons theory}
\date{}
\author{Johan K\"all\'en\footnote{Email address:  \href{mailto:johan.kallen@physics.uu.se}{johan.kallen@physics.uu.se}}
\\ \textit{Department of Physics and Astronomy, Uppsala University}\\
    \textit{ Box 516, 
     SE-75120 Uppsala,
     Sweden}}
\maketitle
\begin{abstract}
\noindent We generalize the framework introduced by Kapustin et al.\ for doing path integral localization in Chern-Simons theory to work on any Seifert manifold. This is done by topologically twisting the supersymmetric theory considered by Kapustin et al., after which the theory takes a cohomological form. We also consider Wilson loops which wrap the fiber directions and compute their expectation values. We discuss the relation with other approaches to exact path integral calculations in Chern-Simons theory.     \end{abstract}

\thispagestyle{fancy}

%\addtocontents{toc}

%\end{titlepage}

\newpage
\tableofcontents

\section{Introduction}
In \cite{Kapustin:2009kz}, Kapustin, Willett and Yaakov introduced a new method to perform exact calculations in Chern-Simons theory on $S^3$. It is based on the method developed by Pestun in \cite{Pestun:2007rz} for performing calculations in supersymmetric Yang-Mills theory on $S^{4}$, and indeed the main motivation of the work \cite{Kapustin:2009kz} is to perform calculations in supersymmetric Chern-Simons theories coupled to matter, theories which are not topological. Nevertheless, setting all the matter fields to zero, the superpartners of the gauge field become auxiliary and the method allows us to calculate the partition function and also special types of Wilson loops in pure Chern-Simons theories on $S^{3}$ in a new way.  
\\
\\
The motivation of this paper is to generalize the method of Kapustin et al.\ to work for a broader class of manifolds. The method of calculation in \cite{Kapustin:2009kz} is localization of the path integral, achieved by introducing a set of auxiliary fields and an auxiliary odd symmetry in the theory. On $S^3$, this symmetry is supersymmetry, and the symmetry transformations are defined using a Killing spinor. It is not straightforward to apply this method as it stands on other types of three manifolds, since the existence of Killing spinors is a non-trivial requirement. Nevertheless, this approach has been attempted in for example \cite{Gang:2009wy} by Gang. Here we choose to attack the problem in a different way. We will perform a variant of topological twisting of the supersymmetry defined on $S^{3}$, and hence introduce an auxiliary odd symmetry in Chern-Simons theory defined without a choice of Killing spinor. Instead, the symmetry is defined using a contact structure and its dual vector field, the Reeb vector field. As shown in \cite{Martinet:1971}, any orientable, compact three manifold admits a contact structure. The introduction of this auxiliary symmetry in Chern-Simons theory, defined on any orientable, compact three manifold with a choice of contact structure, is one of the main results of the present paper. 
\\
\\
The symmetry that we will introduce squares to a gauge transformation plus a translation in the direction of the Reeb vector field, and it acts on the fields in the theory as an equivariant differential. Thus, it is ideal to use for localization of the path integral. However, as we will review below, in order to localize we will need to pick up a metric on the underlying manifold which is invariant under the action of the Reeb vector field. Hence, we need to require that the Reeb vector field generates an isometry. Three manifolds for which this occurs are precisely Seifert manifolds. Seifert manifolds are non-trivial $U(1)$ fibrations over a Riemann surface $\Sigma$, where $\Sigma$ is allowed to have orbifold points. In fact, Chern-Simons theory on Seifert manifolds has an interesting history. Of course, as explained in the foundational paper by Witten \cite{Witten:1988hf}, the Chern-Simons partition function can be computed in the Hamiltonian framework on any three manifold $M$ by exploiting the connection to two dimensional conformal field theory. In \cite{Lawrence:1999lr}, for $M$ a Seifert manifold and $SU(2)$ as the gauge group, Lawrence and Rozansky showed how to manipulate the expression for the partition function obtained with these techniques so that it can be written as a sum of contributions from flat connections. This was later generalized by Mari\~no to any simply-laced group in \cite{Marino:2002fk}. This expression for the partition function has an obvious path integral flavour, and it was later derived from a path integral calculation by Beasley and Witten in \cite{Beasley:2005vf}, using the method of non-abelian localization. Below, we will derive these results using the method based on supersymmetry. Other approaches to Chern-Simons theory on Seifert manifolds includes the one taken by Blau and Thompson in \cite{Blau:2006gh,Thompson:2010iy}, where the method of abelianisation is used, and the one by Jeffrey and McLellan in \cite{Jeffrey:2009vp}, where they consider the case with $U(1)$ as the gauge group.
\\
\\
The most advantageous aspect of the present approach to Chern-Simons theory, as compared to non-abelian localization, is the simplicity of computations of expectation values of Wilson loops. As we will see in section \ref{Wilson section}, the localization computation of a  Wilson loop which wraps the fiber direction of the Seifert manifold is straightforward, and the results agrees completely with the ones obtained in \cite{Beasley:2009mb}, where these types of Wilson loops were considered within the framework of non-abelian localization.
\\
\\
Another interesting aspect of the supersymmetric approach to localization of Chern-Simons theory is the following observation. We will derive that the path integral localizes to the set of equations 
\beq
\begin{split}
F&=0 \\
d_{A}\sigma&=0.
\end{split}
\eeq{}
$F$ is the curvature of the gauge field $A$, $d_{A}$ the gauge covariant derivative and $\sigma$ an auxiliary scalar field in the adjoint representation. For a compact manifold without a boundary, these equations are equivalent to the Bogomolny equations for magnetic monopoles. The appearance of this set of equations as a localization locus sparks the question whether Chern-Simons theory on Seifert manifolds can be handled using the standard machinery of Cohomological topological field theories (TFT's), an approach taken to Chern-Simons theory on $\Sigma\times S^{1}$ by Baulieu, Losev and Nekrasov in \cite{Baulieu:1997nj}. We will address this question in the last part of the paper, and we will see that the approach to Chern-Simons theory considered here is analogous to a generalization of the approach in \cite{Baulieu:1997nj}, for a non-trivial $S^{1}$ fibration over $\Sigma$.   
\\
\\
The paper is organized as follows. In section \ref{background} we review the ingredients which appeared in the localization calculation in \cite{Kapustin:2009kz}, and we will also review the localization method. In sections \ref{Loc Seifert}, \ref{geometric setup} and \ref{gaugefix} we will introduce the twisted model, describe the relevant aspects of Seifert manifolds and also discuss the gauge fixing, which is an important aspect of the present approach. In section \ref{Calc} we will perform a computation of the partition function within the presented framework, and we will find full agreement with the results obtained in \cite{Lawrence:1999lr,Marino:2002fk,Beasley:2005vf}. In section \ref{Wilson section} we discuss the incorporation of Wilson loops wrapping the fiber directions, and how to compute their expectation values. In section \ref{cohTFT} we look at Chern-Simons theory on Seifert manifolds from the perspective of standard Cohomological TFT's, and especially discuss the similarities between the present approach and the one taken in \cite{Baulieu:1997nj}.
\section{Background} \label{background}
We will consider Chern-Simons theory on a compact three manifold $M$ with a compact, simple, simply connected gauge group $G$. Let $P$ be a trivial principal $G$ bundle over $M$, and let $A$ denote a connection on $P$. The Chern-Simons action is defined by 
\beq
S_{CS}=\frac{k}{4\pi}\int_M{\text{Tr}\left(A\wedge dA+\frac{2}{3}A\wedge A\wedge A\right)},
\eeq{}
where $\text{Tr}$ denotes an invariant inner product on $\mathfrak{g}$, the Lie algebra of the gauge group $G$. We will consider the partition function $Z(k)$ defined by the path integral
\beq
Z(k)=\frac{1}{\text{Vol}(\mathcal{G})}\int{\mathcal{D}A~e^{iS_{CS}}}.
\eeq{pathintegral}
Here $k\in\mathbb{Z}$ is the level, and we have divided by the volume of the gauge group $\mathcal{G}$ in the conventional way. With the $"\text{Tr}"$ suitably normalized, $k$ must be integer valued in order for the quantum theory to be gauge invariant. We want to calculate this quantity using localization, a technique which we now describe.
\subsection{Localization} \label{Localization}
The underlying idea of localization is that in some situations the semi-classical approximation to the path integral actually becomes exact. Let us first describe the formula for finite dimensional integrals, where it is called the Atiyah-Bott-Berline-Vergne localization formula \cite{Atiyah:1984,Berline:1982}. The situation where it applies is as follows: Let a group $\mathcal{H}$ act on a manifold $M$ which carries a metric which is $\mathcal{H}$-invariant. Let the action on $M$ be generated by the vector field $v$. We then consider equivariant differential forms, that is, differential forms with values in the Lie algebra of $\mathcal{H}$, invariant under the action of $v$. We form the equivariant differential $Q=d-\phi^{a}i_{v^{a}}$, where $d$ is the de Rham differential, $\phi^{a}$ is a parameter for the action and $i_v$ denotes the contraction of vector fields with differential forms. We have $Q^{2}=-\phi^{a}\mathcal{L} _{v^{a}}$, and hence $Q$ is a differential when acting on equivariant differential forms. Consider now a $Q$-closed equivariant form $\alpha$. The localization formula says that 
\beq
\int_{M}{\alpha}=\int_{F}{\frac{i^{*}_{F}\alpha}{e(N_F)}},
\eeq{cont}
where $F\subset M$ is the set of zeros of the vector field $v$, and $e(N_F)$ is the equivariant Euler class of the normal bundle of $F$. In the case where $F$ is a set of discrete points, the formula reduces to 
\beq
\int_{M}{\alpha}=\sum_{p\in F}{{\frac{i^{*}_{p}\alpha}{\sqrt{\text{det}~ L_p}}}},
\eeq{disc}
where $L_p$ is the action of $\mathcal{H}$ on the tangent space at the point $p$.
\\
\\
This setup can be carried over to path integrals, see for example \cite{Witten:1988ze} and \cite{Witten:1991zz} for the original work. In physics terms, we let $Q$ be an odd symmetry of the action $S$, and we let $Q^2=\mathcal{L}$ be some even symmetry. In our case the even symmetry will be gauge invariance and a certain isometry of the underlying manifold. We are interested in computing the path integral
\beq
Z=\int{e^{iS}}.
\eeq{}
Localization can argued as follows, which mimics the proof of the Atiyah-Bott-Berline-Vergne localization formula in finite dimensions. We can deform the action with a term $tQV$, $t$ some parameter, without changing the path integral, as long as $Q^2 V=0$. This works since
\beq
\frac{d}{dt}Z_t\equiv \frac{d}{dt}\int{e^{iS+tQV}}=\int{QVe^{iS+tQV}}=\int{Q\left(Ve^{iS+tQV}\right)}=0,
\eeq{}
where we have used in the last line that the integral of something $Q$-exact is zero \cite{Witten:1988ze}. Setting $t=0$, we recover the original expression. In order for this to be useful, we need a metric on the space of fields, so that we can construct a positive definite $V$. In this case, letting $t\rightarrow \infty$, the integral will be dominated by the term $tQV$. The only contribution will come from the set of fields at the critical point of QV, and the one-loop approximation becomes exact. Depending on whether the critical point set is discrete or continuous, the path integral is given by the analog of either $\eqref{cont}$ or $\eqref{disc}$, with $\alpha$ given by $e^{iS}$.
\\
\\
As we will see, the space of fields which we will encounter will be a supermanifold, so we should understand the Atiyah-Bott-Berline-Vergne localization formula for supermanifolds. Such a generalization to the super-geometric situation has be considered in for example \cite{Schwarz:1995dg} and \cite{Lavaud}. In this paper, we will restrict our considerations to the case when the fixed point locus is a discrete set of points. In that case, the determinant in $\eqref{disc}$ will be a superdeterminant. In the general case, the equivariant Euler class in the formula $\eqref{cont}$ should be understood in the superformalism. 
\\
\\
We can also include any observable $\mathcal{O}$ in the path integral, and thereby compute its expectation value exactly, as long as $\mathcal{O}$ is preserved by $Q$, that is $Q\mathcal{O}=0$. The natural set of observables in Chern-Simons theory are Wilson loops, and we will discuss the computation of expectation values of Wilson loops in section \ref{Wilson section}.
\\
\\
In order to apply the localization technique, we need an odd symmetry of the action. In the original definition of Chern-Simons theory, no such symmetry exists. We will now describe the way such a symmetry is introduced in \cite{Kapustin:2009kz} when the model is defined on $S^3$. When this is done, we will show how to generalize this framework to work on any Seifert manifold. 

\subsection{Localizing Chern-Simons theory on $S^{3}$}\label{S3}
A way to introduce such a symmetry, without changing the model, is to take the $\mathcal{N}=2$ supersymmetric version of Chern-Simons theory. This version of Chern-Simons theory has been discussed on $\mathbb{R}^{3}$ in \cite{Schwarz:2004yj}, and in \cite{Kapustin:2009kz} it was shown how to generalize the supersymmetry transformations in order to define the theory on $S^{3}$. We will now describe the field content and symmetries of $\mathcal{N}=2$ Chern-Simons theory on $S^{3}$ following \cite{Kapustin:2009kz}, since this will be the starting point for our construction below. The $\mathcal{N}=2$ gauge multiplet consists of the gauge field $A$, two real scalar fields $\sigma$ and $D$, and a 2-component complex spinor $\lambda$. This is a dimensional reduction of the $\mathcal{N}=1$ multiplet in four dimensions. The fields $\sigma, D, \lambda$ are all in the adjoint representation of the gauge group $G$. On $S^3$, the $\mathcal{N}=2$ supersymmetric action is given by 
\beq
S_{\mathcal{N}=2}=\frac{k}{4\pi}\int_{M}{\text{Tr}\left(A\wedge dA+\frac{2}{3}A\wedge A\wedge A\right)}-\frac{k}{4\pi}\int_{M}\text{Tr}{\sqrt{g}\left(\lambda^{\dagger}\lambda-2D\sigma\right)}.
\eeq{s3action}
The supersymmetry transformations are defined using the spinors $\epsilon$ and $\eta$, and are given by
\beq
\begin{split}
\delta A_{\mu}&=\frac{i}{2}\left(\eta^{\dagger}\gamma_{\mu}\lambda-\lambda^{\dagger}\gamma_{\mu}\epsilon\right) \\
\delta\sigma&=-\frac{1}{2}\left(\eta^{\dagger}\lambda+\lambda^{\dagger}\epsilon\right) \\
\delta D&=\frac{i}{2}\left(\eta^{\dagger}\gamma^{\mu}\left(D_{\mu}\lambda\right)-\left(D_\mu \lambda^{\dagger}\right)\gamma^{\mu}\epsilon\right)+\frac{i}{6}\left(\nabla_{\mu}\eta^{\dagger}\gamma^{\mu}\lambda-\lambda^{\dagger}\gamma^{\mu}\nabla_{\mu}\epsilon\right)-\frac{i}{2}\left(\eta^{\dagger}[\lambda,\sigma]-[\lambda^{\dagger},\sigma]\epsilon\right) \\
\delta\lambda &=\left(-\frac{1}{2}\gamma^{\mu\nu}F_{\mu\nu}-D+i\gamma^{\mu}D_{\mu}\sigma\right)\epsilon +\frac{2i}{3}\sigma\gamma^{\mu}\nabla_{\mu}\epsilon \\
\delta \lambda^{\dagger}&=\eta^{\dagger}\left(\frac{1}{2}\gamma^{\mu\nu}F_{\mu\nu}-D-i\gamma^{\mu}D_{\mu}\sigma\right)-\frac{2i}{3}\sigma\nabla_{\mu}\eta^{\dagger}\gamma^{\mu}.
\end{split}
\eeq{susy}
Above, $g$ is the determinant of the metric on $S^3$, $\gamma_{\mu}$ are the Pauli matrices, which are hermitian, and we have defined $\gamma_{\mu\nu}\equiv\frac{1}{2}\left(\gamma_{\mu}\gamma_{\nu}-\gamma_{\nu}\gamma_{\mu}\right)$. The derivative $D_{\mu}$ is covariant both with respect to the gauge field and the spin connection: $D_{\mu}=\nabla_{\mu}+[A,~]$, where $[~,~]$ is the Lie algebra bracket and $\nabla_{\mu}$ is the spinor covariant derivative. Finally, $F_{\mu\nu}$ is the field strength, defined by $F=dA+A\wedge A$. 
\\
\\
We see that the scalar and spinor fields can be integrated out, and hence
\beq
Z(k)=\frac{1}{\text{vol}(\mathcal{G})}\int{\mathcal{D}A~e^{iS_{CS}}}=\frac{1}{\text{vol}(\mathcal{G})}\int{\mathcal{D}A\mathcal{D}\lambda\mathcal{D}\sigma\mathcal{D}D~e^{iS_{\mathcal{N}=2}}}.
\eeq{}
The advantage of keeping the auxiliary fields is that now we have the sought fore odd symmetry at hand, namely the supersymmetry. 
\\
\\
In order to localize, we need to pick a specific supersymmetry. In \cite{Kapustin:2009kz}, the transformation defined by
\beq
\begin{split}
\nabla_{\mu}\epsilon&=\frac{i}{2}\gamma_{\mu}\epsilon, \quad \quad \quad \epsilon^{\dagger}\epsilon=1 \\
\eta&=0
\end{split}
\eeq{killingspinor}
is chosen. The first equation obeyed by $\epsilon$ shows that it is a Killing spinor. With these values of $\epsilon$ and $\eta$, the transformations $\eqref{susy}$ fulfill $\delta^{2}=0$ on all fields, and it can be used to localize the path integral. 
\\
\\
We will now generalize the above construction to work on more general manifolds. 
\section{Localizing Chern-Simons theory on Seifert manifolds} \label{Loc Seifert}
The odd symmetry used in \cite{Kapustin:2009kz} to localize is defined using a Killing spinor $\epsilon$. Moreover, the auxiliary field $\lambda$ is also a spinor. Searching for Killing spinors on more general three manifolds than $S^{3}$ is not something we want to do. For example, not all Lens spaces admits Killing spinors \cite{Franc:1987}. Instead, we will redefine the spinors into differential forms, a procedure known as topological twisting. The twisting is done using a geometrical structure which any compact, orientable three manifold posses, namely a contact structure \cite{Martinet:1971}. A contact structure is a one-form $\kappa$ such that $\kappa\wedge d\kappa\neq 0$. $\kappa\wedge d\kappa$ can therefore serve as a volume form. Given a contact structure, there always exists a vector field $v$ such that 
\beq
i_v\kappa=1, \quad \quad i_v d\kappa=0.
\eeq{reeb}
$v$ is known as the Reeb vector field. For more information about contact manifolds, see for example the book \cite{Blair:2010}.
\\
\\
Since the end result after the twist is performed has a nice and simple form, we will just state it here, leaving the derivation to appendix \ref{twist}. The spinors $\lambda$ and $\lambda^{\dagger}$ appearing in the model defined on $S^{3}$ are redefined into an odd zero-form $\alpha$ and an odd one-form $\Psi$, both with values in the Lie algebra $\mathfrak{g}$ of the gauge group $G$: $\alpha\in\Omega^{0}(M,\mathfrak{g})$ and $\Psi\in\Omega^{1}(M,\mathfrak{g})$. Nothing happens to the even scalars $\sigma,D$, which we remind ourselves both takes values in $\mathfrak{g}$. Given a contact structure $\kappa$, we can write down the action\footnote{We hope that there will arise no confusion between the Chern-Simons level $k$ and the contact structure $\kappa$. Both notations are standard in the literature, and therefore kept here.}
\beq 
\begin{split}
S&=\frac{k}{4\pi}\int_M{\text{Tr}\left(A\wedge dA+\frac{2}{3}A\wedge A\wedge A-~\kappa\wedge \Psi\wedge \Psi-2d\kappa\wedge\Psi\alpha+\kappa\wedge d\kappa~D\sigma\right)} \\
&\equiv S_{CS}+S_{aux}.
\end{split}
\eeq{twistaction}
As before, all fields except the gauge field $A$ are auxiliary and can be integrated out. The following transformations is a symmetry of the above action:
\beq
\begin{split}
\delta A&=\Psi \\
\delta \Psi&=i_v F+id_{A}\sigma \\
\delta \alpha &= -\frac{\kappa\wedge F}{\kappa\wedge d\kappa}+\frac{i}{2}\left(\sigma+D\right)\\
\delta\sigma&=-i~i_v\Psi \\
\delta D&=-2i\mathcal{L}^{A}_v\alpha-2[\sigma,\alpha]-2i\frac{\kappa\wedge d_{A}\Psi}{\kappa\wedge d\kappa}+i~i_v\Psi.
\end{split}
\eeq{auxsym}
Here $v$ denotes the Reeb vector field, $d_A=d+[A,~]$ is the de Rham differential twisted by the gauge field, $\mathcal{L}_v=di_v+i_vd$ is the Lie derivative in the direction of the vector field $v$ and we have defined $\mathcal{L}^{A}_v=\mathcal{L}_v+[i_vA,~]$. The notation $\frac{1}{\kappa\wedge d\kappa}$ is inspired from \cite{Beasley:2005vf} should be understood as follows. Since $\kappa\wedge d\kappa$ is non-vanishing, any 3-form, for example $\kappa\wedge F$, can be written as $\kappa\wedge F=\kappa\wedge d\kappa ~ \gamma$, for some $\gamma\in\Omega^{0}(M,\mathfrak{g})$. By $\frac{\kappa\wedge F}{\kappa\wedge d\kappa}$ we mean this function $\gamma$. 
\\
\\
The transformations above fulfill $\delta^{2}=\mathcal{L}_v+G_{\Phi}$, where $G_{\Phi}$ denotes a gauge transformation with parameter $\Phi=i\sigma-i_v A$. The gauge transformations acts as $G_{\Phi}A=d_A \Phi$ on the gauge field and as $G_{\Phi} X=-[\Phi,X]$ on any other field. The situation here is different from the one in \cite{Kapustin:2009kz}, where they used a symmetry which squared to zero. The difference is due to a slightly different choice of spinors in equation $\eqref{susy}$. See appendix \ref{twist} for the details on how to obtain $\eqref{twistaction}$ and $\eqref{auxsym}$ starting from $\eqref{s3action}$ and $\eqref{susy}$.
\\
\\
The transformations $\eqref{auxsym}$ can be defined on any orientable, compact three manifold, with a choice of contact structure. But this is not enough to localize. As explained in section \ref{Localization}, we also have to introduce a positive definite invariant inner product on the tangent space of the space of fields. One way of doing this is by choosing a metric on $M$ and using the trace on the Lie algebra. Given two tangent vectors $X$ and $Y$, we define their inner product by
\beq
\left(X,Y\right)=\text{Tr}\int_M{X\wedge *Y},
\eeq{metric}
where $*$ is the Hodge star defined using the metric on $M$. Following section \ref{Localization}, to localize we will choose\footnote{The bar denotes complex conjugation, introduced since $\delta$ is a complex transformation. See the discussion at the end of this subsection of how to think about the complex transformation in this context. The bar does nothing to the Lie algebra generators.}  $V=\left(\Psi,\overline{\delta\Psi}\right)+\left(\alpha,\overline{\delta \alpha}\right)$, since the bosonic part of $\delta V$ is then positive definite. We need to make sure that $\delta ^{2} V=0$. A straightforward computation gives
\beq
\begin{split}
\delta ^{2} V= &\text{Tr}\int_M{\mathcal{L}_v\Psi\wedge *\overline{\delta\Psi}}+\text{Tr}\int_M{\Psi\wedge *\mathcal{L}_v\overline{\delta\Psi}}+\text{Tr}\int_M{G_{\Phi}\Psi\wedge *\overline{\delta\Psi}}+\text{Tr}\int_M{\Psi\wedge *G_{\Phi}\overline{\delta\Psi}} \\
&+ \text{same terms with }\alpha.
\end{split}
\eeq{}
All fields transform in the adjoint representation of the gauge group, so the last two terms cancel since the trace is cyclic. For the first two terms, we need to require that the vector field $v$ is a Killing vector field in order to be able to pull the derivative through the Hodge star and perform a partial integration to obtain cancelation. That is, we have to require that the Reeb vector field generates an isometry on $M$. Compact, orientable three manifolds with this property are precisely Seifert manifolds (see for example theorem 7.1.3 in \cite{Boyer:2008}). Hence, our localization framework works on Seifert manifolds. 
\\
\\
We will give a quick description of the aspects of Seifert manifolds which are needed in the present work in section \ref{geometric setup}. Before doing so, we will determine to which subspace the Chern-Simons path integral localizes, and we will also write the transformations $\eqref{auxsym}$ in a more geometrical form via a change of variables.
\\
\\
Before continuing, we will address a slight subtlety with the transformations defined in $\eqref{auxsym}$, in connection to the localization framework. All our fields in the theory are real, and one could worry about the fact that the transformations $\eqref{auxsym}$ are complex; the transformed field is not real anymore. The same issue appears in \cite{Pestun:2007rz}, where it is resolved in the following way. We can understand the action $\eqref{twistaction}$ as analytically continued to the space of complexified fields. The path integral is understood as an integral of a holomorphic functional over a half-dimensional contour of integration in the space of complexified fields. As for the complex conjugation appearing in $V$ above, it should be understood as complex conjugation when the path integral is restricted to the original integration contour, namely the one over the "real axis". For a general contour of integration, we choose a $V$ where $\overline{\delta\Psi}$ and $\overline{\delta \alpha}$ are \textit{defined} by switching the signs of the factors of $"i"$ on the right hand side in equation $\eqref{auxsym}$. When the contour of integration is restricted to the real axis, the bosonic part of $V$ is positive definite, and the localization argument goes through.
\subsection{Fixed point locus} \label{fixed point}
The field configurations to which the theory localizes is given by the fixed point set of the transformations $\eqref{auxsym}$ on the fermionic fields. With the above understanding of how to think of the complex transformation $\delta$ and the localization, its real and imaginary parts has to vanish separately. From $\eqref{auxsym}$ we find that the fixed point set is given by the equations
\beq
\begin{split}
F&=0 \\
d_{A}\sigma&=0 \\
\sigma+D&=0.
\end{split}
\eeq{F}
We see that the theory localizes to the space of flat connections and covariantly constant scalar fields. This is the same set as found on $S^3$ in \cite{Kapustin:2009kz}. It is interesting to notice that the first two equations are the equivalent of the Bogomolny equations for the manifolds considered here, namely compact manifolds with no boundary. The appearance of this equation in the present treatment of Chern-Simons theory suggests that there should be a relation to the approach taken in \cite{Baulieu:1997nj}. In that work, Chern-Simons theory on $\Sigma\times S^{1}$, where $\Sigma$ is a Riemann surface, is constructed using the framework of Cohomological topological field theory, and the above equations mark their starting point. We will further comment on this aspect of the present work in section \ref{cohTFT}.  
\subsection{Cohomological form} \label{coh1}
Our fields are integration variables in the path integral, and by a change of variables we can write the transformations $\eqref{auxsym}$ in a more suggestive form. Let us change variables from $\sigma, D$ to new variables $\Phi, \tilde{D}$, defined by
\beq
\begin{split}
\Phi&=i\sigma-i_v A \\
\tilde{D}&=-\frac{\kappa\wedge\ F}{\kappa\wedge d\kappa}+\frac{i}{2}\left(\sigma+D\right).
\end{split}
\eeq{cov}
Written in terms of the new variables, the transformations $\eqref{auxsym}$ take the nice cohomological form
\begin{align} \label{cohom}
\delta A&=\Psi, &  \delta\alpha&=\tilde{D}\nonumber \\
\delta\Psi&=\mathcal{L}_v A +G_{\Phi}A,  &\delta \tilde{D}&=\mathcal{L}_v \alpha +G_{\Phi}\alpha \\
\delta\Phi&=0\nonumber.
\end{align}
The fields $A,\alpha$ can now be interpreted as coordinates on an infinite dimensional supermanifold. The transformation $\delta$ is acting as an equivariant differential, where the group $\mathcal{H}$ acting on the manifold is given by the semi-direct product $ U(1)\ltimes\mathcal{G}$. The fields $\Psi,\tilde{D}$ can be interpreted as the de Rham differentials of the coordinates $A,\alpha$, whereas the field $\Phi$ is the parameter for the gauge transformations. 
\section{Basic facts of Seifert manifolds}  \label{geometric setup}
As found out in the previous section, the localization framework constructed above can be applied to Chern-Simons theory defined on Seifert manifolds. We will now give some basic facts about these manifolds, closely following the description found in section 3 in \cite{Beasley:2005vf}. For a more complete reference of Seifert manifolds we refer to \cite{Orlik} and the book \cite{Boyer:2008}.
\\
\\
A Seifert manifold is a $U(1)$ bundle over an orbifold $\hat{\Sigma}$:  
\beq
U(1)\rightarrow M\xrightarrow{\pi} \hat{\Sigma}.
\eeq{}
It is described by the integers
\beq
\begin{split}
&\left[g;n;(a_1,b_1), (a_2,b_2)\ldots (a_N,b_N)\right] \\
&\text{gcd}(a_j,b_j  )=1,\quad \quad 0<b_j  <a_j.
 \end{split}
\eeq{seifertinv}
For completeness, let us briefly describe this data. $\hat{\Sigma}$ is a Riemann surface of genus $g$ with $N$ marked points $p_j$, $j=1,2 \ldots N$. The neighbourhood of the points $p_j$ is modelled not by $\mathbb{C}$ but $\mathbb{C}/\mathbb{Z}_{a_j }$, where $\mathbb{Z}_{a_j}$ is a cyclic group acting on the local coordinates $z$ at $p_j$ as
\beq
z\mapsto \xi\cdot z,   \quad \quad \xi=e^{\frac{2\pi i}{a_j }}.
\eeq{}
Where the points $p_{j}$ are is topologically irrelevant, and $\hat{\Sigma}$ is completely specified by $g$ and $\{a_{1},a_{2}\ldots a_{N}\}$.
\\
\\
Now we consider a line bundle over $\hat{\Sigma}$, which we denote by\footnote{Note that we denote the line bundle by $\mathcal{L}$ and the Lie derivative along the vector $v$ by $\mathcal{L}_v$. We hope that it is clear from the context what is meant, and that no confusion will arise.}  $\mathcal{L}$. The local trivialization over each orbifold point $p_j$ is now modeled as $\left(\mathbb{C}\times\mathbb{C}\right)/\mathbb{Z}_{a_j}$, where $\mathbb{Z}_{a_j}$ acts on the local coordinates $(z,s)$ of the base and the fiber as
\beq
z\mapsto \xi\cdot z, \quad \quad s\mapsto \xi^{b_j  }\cdot s, \quad\quad  \xi=e^{\frac{2\pi i}{a_j }}
\eeq{}
for some integers $0<b_j  <a_j$. An arbitrary Seifert manifold can be described as the total space of the associated $S^1$ fibration. In order to have a smooth manifold, the group action has to be free. That is why we require $b_j  \neq 0$, and in addition we require each pair of integers $(a_j,b_j  )$ to be relatively prime. Lastly, once we fix the integers $\{b_1,b_2\ldots b_N\}$, the line bundle over $\hat{\Sigma}$ is describe by the integer $n$, the degree. That concludes the description of the data entering $\eqref{seifertinv}$.
\\
\\
On a Seifert manifold, we can choose the contact structure $\kappa$ to be the connection, and therefore $d\kappa$ is the curvature. The Reeb vector field $v$ generates the $U(1)$ rotations of the fibers. As explained in section 3 of \cite{Beasley:2005vf}, the integral of $\kappa\wedge d\kappa$ over $M$ is given in terms of the Seifert invariants described above:
\beq
\int_M{\kappa\wedge d\kappa=n+\sum_{j=1}^{N}{\frac{b_{j}}{a_j}} }.
\eeq{vol}
\subsection{Seifert homology spheres} \label{seiferthomolgy}
Flat connections on a manifold $M$ can be described as the space of group homomorphism from the fundamental group $\pi_1(M)$ to the gauge group $G$, modulo conjugation with elements in $G$. We will in the computations to be performed below focus on the situation when the trivial connection is an isolated point in the space of flat connections, and compute the contribution to the partition function arising from this point in the space of solutions to $\eqref{F}$. The contribution from other parts of the solution space can also be treated within the present approach, and we will comment about these other situations further in the discussion section. The trivial connection is isolated if the first homology group of $M$ vanishes:
\beq
H_{1}(M,\mathbb{R})=0,
\eeq{}
that is, when $M$ is a rational Seifert homology sphere. The conditions on the Seifert invariants for this to occur is the following. Let $d=\left|H_{1}(M,\mathbb{Z})\right|$ and $P=\prod_{j=1}^{N}{a_j }$. Then $H_{1}(M,\mathbb{R})=0$ if 
\beq
\begin{split}
g&=0 \\
\frac{d}{P}&=n+\sum_{j=1}^{N}{\frac{b_j  }{a_j}} \\
\text{gcd}(a_i,a_j)&=1, \quad i\neq j.
\end{split}
\eeq{h10}                           
For an explanation of how these conditions occur, we refer to \cite{Beasley:2005vf}.
\\
\\
From now on we will restrict our considerations to manifolds whose Seifert invariants fulfill the conditions $\eqref{h10}$.                               
\section{Gauge fixing} \label{gaugefix}
Before we can localize, we have to take into account that the action $\eqref{twistaction}$ has gauge invariance, and we only want to integrate over gauge equivalence classes in the path integral $\eqref{pathintegral}$. In the standard way, this is achieved by introducing ghost fields along with a BRST transformation $\delta_B$. As in \cite{Pestun:2007rz}, we have to make sure that the gauge fixing using this BRST symmetry is compatible with the localization framework constructed above. Specifically, to localize we need to use a transformation $Q$ which is a symmetry of the \textit{gauge fixed} action, which is typically given by
\beq
S_{g.f.}=S+\delta_{B}V_{g.f.}.
\eeq{}
Here $V_{g.f.}$ is some gauge fixing term, and $S$ is our original action in $\eqref{twistaction}$. Unfortunately, the transformation $\eqref{auxsym}$ (with $\delta$ acting trivially on the ghost fields) is not a symmetry of $S_{g.f.}$, something we need to require in order for localization to work. This is easily fixed. When localizing, we will instead use the combined symmetry $Q=\delta + \delta_{B}$, together with a gauge fixing term $QV_{g.f.}$. 
\\
\\
Thus, in order to get a well defined path integral, we will extend the supermanifold of fields to include the set of ghost fields, in addition to the physical fields (that is, the gauge field and auxiliary fields which we considered in section \ref{Loc Seifert}). As we will find out below, the fixed points of the symmetry $Q$ will not change the localization locus for the physical fields, the path integral will still be localized to an integration over the space of flat connections and covariantly constant scalar fields. However, we will find that $Q^{2}=\mathcal{L}_v+G_{a_{0}}$, where $a_{0}$ is a \textit{constant} field (to be introduced below). Therefore, the group $\mathcal{H}$ which acts on our supermanifold is now $\mathcal{H}=G\times U(1)$, where $G$ is the group of constant gauge transformations. 
\\
\\
Following \cite{Pestun:2007rz}, we introduce the set of ghost fields $c,\bar{c},b,a_0,\bar{a}_0,\bar{c}_0,b_0,c_0$, along with the BRST transformations
\begin{align}
\delta_B A&=d_A c \nonumber\\
 \delta_B\Phi&= -[c,\Phi]-\mathcal{L}_v c \nonumber\\
 \delta_B X&=-[c,X] \\
\delta_B c&=a_0-\frac{1}{2}[c,c], & \delta_B\bar{c}&=b, & \delta_B\bar{a}_0&=\bar{c}_0, & \delta_B b_0&=c_0 \nonumber \\
\delta_B a_0&=0, & \delta_B b&=-[a_0,\bar{c}], & \delta_B \bar{c}_0&=-[a_0,\bar{a}_0], & \delta_B c_{0}&=-[a_0,b_0].\nonumber
\end{align}
As usual, $X$ denotes all the fields which do not appear explicitly above. The transformation of $\Phi$ is derived from its definition in $\eqref{cov}$. The odd fields $c\in\Omega^{0}(M,\mathfrak{g})$,  $\bar{c}\in\Omega^{3}(M,\mathfrak{g})$ and the even field $b\in\Omega^{3}(M,\mathfrak{g})$ are the standard Faddeev-Popov ghosts. The fields $a_0,\bar{a}_0,\bar{c}_0,b_0,c_0$ are introduced in order to treat the zero modes of the fields $c,\bar{c},b$ in a systematic way. They are all constant fields, $a_0,\bar{a}_0,b_0$ are bosonic whereas $c_0,\bar{c}_0$ are fermionic. With the inclusion of this set of constant fields, the BRST operator $\delta_B$ squares to a constant gauge transformation generated by $a_0$:
\beq
\delta^{2}_B=G_{a_0}.
\eeq{}
The transformations $\delta_B$ is a symmetry of our original action $\eqref{twistaction}$. We will now combine this BRST symmetry with the symmetry $\delta$ defined in $\eqref{auxsym}$, thereby defining a new symmetry $Q=\delta+\delta_B$. This will be the symmetry which we will use to localize. 
\subsection{Combining the supersymmetry with the BRST symmetry}
To simplify the formulas, let us introduce a collective notation for the physical fields appearing in $\eqref{cohom}$:
\beq
Z=\left\{A,\alpha\right\}, \quad\quad Z'=\{\Psi,\tilde{D}\}.
\eeq{}
If we define the $\delta$-transformations for the ghost fields in a judicious way, we find the action of $\delta$ on all our fields to be
\begin{align} \label{aux+brst}
\delta Z&=Z' \nonumber\\
\delta Z&'=\mathcal{L}_vZ+G_{\Phi}Z\nonumber \\
\delta a_0&=0 \\
\delta c&=-\Phi, & \delta \bar{c}&=0, & \delta \bar{a}_0&=0, & \delta b_0&=0 \nonumber\\ 
\delta \Phi&=0, & \delta b&=\mathcal{L}_v\bar{c}, & \delta \bar{c_0}&=0, & \delta c_0&=0. \nonumber\\
\end{align}
We now want to combine these transformations with the $\delta_B$-transformations, which are given by
\begin{align}
\delta_B Z&=G_{c}Z \nonumber\\
\delta_B Z'&=G_{c}Z' \nonumber\\
\delta_B a_0&=0\\
\delta_B c&=a_0-\frac{1}{2}[c,c], & \delta_B \bar{c}&=b, & \delta_B \bar{a}_0&=\bar{c}_0, & \delta_B b_0&=c_0 \nonumber\\ 
\delta_B \Phi&=-\mathcal{L}_vc-[c, \Phi], & \delta_B b&=-[a_0,\bar{c}], & \delta_B \bar{c}_0&=-[a_0,\bar{a}_0], &Ê\delta_B c_0&=-[a_0,b_0]. \nonumber
\end{align}
We simply define a new operator 
\beq
Q=\delta+\delta_B.
\eeq{}
The action of $Q$ on all fields are given by
\begin{align} \label{Q}
QZ&=Z'+G_cZ \nonumber \\
QZ'&=\mathcal{L}_vZ+G_{\Phi}Z+G_cZ' \nonumber\\
Qa_0&=0 \\
Qc&=a_0-\Phi-\frac{1}{2}[c,c], & Q\bar{c}&=b, & Q\bar{a}_0&=\bar{c}_{0}, & Qb_0&=c_0\nonumber \\
Q\Phi&=-\mathcal{L}_vc+G_{c}\Phi, & Qb&=\mathcal{L}_v\bar{c}+G_{a_0}\bar{c}, & Q\bar{c}_0&=G_{a_0}\bar{a}_0, & Qc_0&=G_{a_0}b_0. \nonumber
\end{align}
Using the above relations, we find that the operator $Q$ squares to a Lie derivative in the $v$-direction plus a gauge transformation with constant parameter $a_0$: 
\beq
Q^2=\mathcal{L}_v+G_{a_0}.
\eeq{}
\subsection{Integrating over gauge equivalence classes}
The simplest way to only integrate over gauge equivalence classes in the the path integral $\eqref{pathintegral}$ is to add a term to the action $\eqref{twistaction}$ which restricts the integration of the gauge field $A$ to be transversal to the gauge orbits. We will choose to work in the Lorenz gauge, which is defined by
\beq
d^{\dagger}A=0,
\eeq{}
where $d^{\dagger}$ is the adjoint operator to $d$ using the inner product defined by the Hodge star. Adding the term $iQV_{g.f.}$, with 
\beq
V_{g.f.}=(\bar{c},d^{\dagger}A)+(\bar{c},b_{0})+(c,\bar{a}_0),
\eeq{}
will restrict the integration to field configurations which satisfies the Lorenz gauge, as we now demonstrate. Written out explicitly, $QV_{g.f}$ is given by
\beq
QV_{g.f.}\equiv S_{g.f.}=(b,d^{\dagger}A)-(\bar{c},d^{\dagger}d_{A}c)-(\bar{c},d^{\dagger}\Psi)+(b,b_0)-(\bar{c},c_0)+(a_0-\Phi-\frac{1}{2}[c,c],\bar{a}_0)-(c,\bar{c}_0).
\eeq{QV}
Here, the first two terms are the standard ones. The zero modes of $c,\bar{c}$ and $b$ do not appear in these two terms, so these zero modes, together with $\bar{a}_0$, can be used to integrate out the last four terms. As discussed in detail in \cite{Pestun:2007rz}, the term $(\bar{c},d^{\dagger}\Psi)$ can be absorbed in the second one using a change of variables, with trivial Jacobian. We are therefore left with the integration over $c,\bar{c},b$, without the zero modes, and the two terms
\beq
(b,d^{\dagger}A)-(\bar{c},d^{\dagger}d_{A}c).
\eeq{}
The first term restricts the integration to the surface $d^{\dagger}A=0$, whereas the second one arises from the Faddeev-Popov determinant. In summary, the gauge fixing is done in the following way:
\beq
\begin{split}
Z&=\frac{1}{\text{Vol}(\mathcal{G})}\int{\mathcal{D}Xe^{iS[X]}}=\frac{1}{\text{Vol}(\mathcal{G})}\int{\mathcal{D}Xe^{iS[X]}}\int_{g\in \mathcal{G}'}{\mathcal{D}g~\delta(d^{\dagger}A^{g})~\text{det}'(d^{\dagger}d_A)} \\
&=\frac{\text{Vol}(\mathcal{G}')}{\text{Vol}(\mathcal{G})}\int{\mathcal{D}X\mathcal{D}b'\mathcal{D}\bar{c}'\mathcal{D}c'e^{iS[X]+i(b,d^{\dagger}A)-i(\bar{c},d^{\dagger}d_A c)}} \\
&=\frac{1}{\text{Vol}(G)}\int{\mathcal{D}X\mathcal{D}b\mathcal{D}b_0\mathcal{D}\bar{c}\mathcal{D}\bar{c}_0\mathcal{D}c\mathcal{D}c_0\mathcal{D}\bar{a}_0\mathcal{D}a_0e^{iS[X]+iQV_{g.f.}}}.
\end{split}
\eeq{Zgf}
Above, $X$ denotes all physical fields, $'$ denotes the exclusion of the zero modes, $A^{g}$ denotes the gauge transformed connection $A$, and $\mathcal{G}'=\mathcal{G}/G$ is the coset of the group of gauge transformations by constant gauge transformations. We notice that in the last line the integration over the fields $b,c,\bar{c}$ \textit{includes} the zero modes.
\subsection{Cohomological form of the Q-complex} \label{coh2}
Let us make yet another change of variables, in order to put the $Q$ transformations defined in $\eqref{Q}$ into a cohomological form. We let 
\beq
\begin{split}
\tilde{Z}'&=Z'+G_{c}Z \\
\tilde{\Phi}&=a_0-\Phi-\frac{1}{2}[c,c].
\end{split}
\eeq{}
Then the transformations $\eqref{Q}$ are written as
\begin{align} \label{finalQ}
QY&=Y', & QY'=\mathcal{L}_vY+G_{a_{0}}Y \\
Qa_0&=0 \nonumber
\end{align}
where $Y=\{Z,c,\bar{c},\bar{a}_0,b_0\}$ and $Y'=\{\tilde{Z}',\tilde{\Phi},b,\bar{c}_0,c_0\}$. We have now extended the supermanifold encountered in section \ref{coh1}, and included the ghost fields as coordinates. $Y$ is interpreted as a coordinate on an infinite dimensional supermanifold $\mathcal{M}$. $Y'$ is the de Rham differential of $Y$. We have a group $\mathcal{H}=G\times U(1)$ acting on $\mathcal{M}$. $Q$ can be thought of as the equivariant differential with respect to this group action, and the constant field $a_0$ is a parameter for the action of the group $G$. 
 \subsection{Summary of the localization procedure}
We now have all ingredients needed in order to perform the localization of the Chern-Simons partition function on Seifert manifolds $M$. Let us summarize. We have introduced the auxiliary fields $\Psi,\alpha,\sigma,D$ together with gauge fixing fields $c,\bar{c},b,a_0,\bar{a}_0,\bar{c}_0,b_0,c_0$. We have also introduced the odd transformations $Q$ defined in equation $\eqref{Q}$. $Q$ is a symmetry of the action
\beq
\tilde{S}=S_{CS}+S_{aux}+S_{g.f.},
\eeq{Stot}
and fulfills $Q^2=\mathcal{L}_v+G_{a_0}$. Here $v$ is the Reeb vector field which generates an isometry on $M$, and $G_{a_0}$ is a gauge transformation generated by the constant field $a_0$. The path integral over all fields $A,\Psi,\alpha,\sigma,D,c,\bar{c},b,a_0,\bar{a}_0,\bar{c}_0,b_0,c_0$ with the action $\tilde{S}$ gives the same result as the path integral over the gauge field with the Chern-Simons action alone, restricted to the gauge fixing surface $d^{\dagger}A=0$. By a series of redefinitions of our fields, the $Q$ transformations can be put in the form $\eqref{finalQ}$, and hence we can apply the localization procedure as explained in section \ref{Localization}.  The group which we use to localize is given by 
\beq
\mathcal{H}=G\times U(1).
\eeq{}
The localization locus is determined by looking at the fixed points of the $Q$ action. For the physical fields, we determined the fixed point in $\eqref{F}$, and this locus remains the same in the gauge fixed theory, with the additional constraint that the zero mode of $\Phi$ is identified with $a_0$, as can be seen from $\eqref{QV}$\footnote{In order to make this identification we should integrate the \textit{zero mode} of $\sigma$ over the imaginary axis, so that the zero mode of $\Phi$ becomes purely real, see equation $\eqref{cov}$.}.
\section{Calculation of the partition function} \label{Calc}
We will now calculate the partition function when $M$ is a Seifert homology sphere, focusing on the contribution which arise from the point in the space determined by $\eqref{F}$ which corresponds to the trivial connection. The solution to the equations in $\eqref{F}$ then reads
\beq
\begin{split}
A&=0 \\
\text{constant}&=\sigma=-D\equiv \phi, 
\end{split}
\eeq{trivialfp}
where we denoted the zero mode of $\sigma$ (and hence $\Phi$ in this case) by $\phi$.  For a Seifert homology sphere, the trivial connection is an isolated point, and the path integral $\eqref{Zgf}$ is given by the analog of the right hand side of $\eqref{disc}$, together with the residual integration over $\phi$. Since we are on a supermanifold, the determinant $\frac{1}{\sqrt{\det~L}}$ in the expression $\eqref{disc}$ should be understood as a superdeterminant. It is the determinant of the operator $L_{a_0}=\mathcal{L}_v+[a_0,~]$, with respect to the action on the tangent space of the space of fields. Therefore, it is a function of $a_0$. Using the constraint $a_0=\phi$, the determinant $\frac{1}{\sqrt{\det~L}}$ becomes a function of $\phi$, and from now on we will denote it by $h(\phi)$.
\\
\\
We will now evaluate the action $\eqref{twistaction}$ on the solution $\eqref{trivialfp}$. Using equations $\eqref{vol}$ and $\eqref{h10}$, we get
\beq
\begin{split}
\frac{k}{4\pi}\int_{M}{\kappa\wedge d\kappa ~\text{Tr}(\phi^{2})}=\frac{1}{2\epsilon}\left(\frac{d}{P}\right)~\text{Tr}(\phi^{2}),
\end{split}
\eeq{}
where we have defined
\beq
\epsilon\equiv\frac{2\pi}{k}.
\eeq{}
Putting all the pieces together, we find that the contribution to  the path integral $\eqref{Zgf}$ which comes from the point corresponding to the trivial connection will be given by the
\beq
Z_{\{0\}}=\frac{1}{\text{Vol}(G)}\int_{\mathfrak{g}}{[d\phi] ~\exp{\left[\frac{i}{2\epsilon}\left(\frac{d}{P}\right)\text{Tr}\left(\phi^{2}\right)\right]}h(\phi)}. \\
\eeq{Zres}
We can further simplify the above formula by exploiting the gauge invariance of the integrand to reduce the integral over the Lie algebra $\mathfrak{g}$ to an integral over the Cartan subalgebra $\mathbf{t}$. This can be done using the Weyl integral formula, which states
\beq
\int_{\mathfrak{g}}{[d\phi]f(\phi)}=\frac{1}{\left|W\right |}\frac{\text{Vol}(G)}{\text{Vol}(T)}\int_{\mathbf{t}}{~[d\phi]~\prod_{\beta>0}{\langle\beta,\phi\rangle^{2}}~f(\phi)}
\eeq{}
for a function $f$ which is invariant under the adjoint action of $\mathfrak{g}$. Above, $|W|$ is the order of the Weyl group of $G$, $T$ is the maximal torus of $G$ and $\beta$ denotes the roots of $\mathfrak{g}$. Therefore, our integral $\eqref{Zres}$ can be written
\beq
Z=\frac{1}{\left|W\right |}\frac{1}{\text{Vol}(T)}\int_{\mathbf{t}}{[d\phi]~ \prod_{\beta>0}{\langle\beta,\phi\rangle^{2}}~\exp{\left[\frac{i}{2\epsilon}\left(\frac{d}{P}\right)\text{Tr}\left(\phi^{2}\right)\right]}h(\phi)}.
\eeq{Zres2}
We now turn to the computation of $h(\phi)$. As we will see, the calculation of $h(\phi)$ is quite analogous the ones performed in \cite{Beasley:2005vf} and \cite{Blau:2006gh}. 
\\
\\
Firstly, there is a possibility of a non-trivial phase, and we write
\beq
h(\phi)=e^{-\frac{i\pi}{2}\eta}\left| h(\phi)\right|.
\eeq{}
Here $\eta$ represents the phase of $h(\phi)$, which needs to be regularized. This phase will give rise to the shift in the Chern-Simons level, and it occurs since the operator $iL_\phi$ has both positive and negative eigenvalues. Schematically it can be written as:
\beq
\eta=\frac{1}{2}\sum_{\lambda}{\text{sign}(\lambda)},
\eeq{eta}
where $\lambda$ are the eigenvalues of $iL_{\phi}$. The above expression is not well defined as it stands, and we will define a regularized version of it when we consider the technical details of the computation in appendix \ref{Details}. For now, we focus on the absolute value of $h(\phi)$. 
\\
\\
The fields which are coordinates on the supermanifold $\mathcal{M}$ are given by
\beq
\text{Bosonic}:\{A,\bar{a}_0,b_0\} \quad\quad \text{ Fermionic}: \{\alpha,c,\bar{c}\}.
\eeq{}
Let $\mathcal{A}$ denote the subspace of $\mathcal{M}$ which contains the gauge field $A$. The tangent space to $\mathcal{A}$ at any point $A_0\in\mathcal{A}$ is isomorphic to the space of one-forms with values in $\mathfrak{g}$:
\beq
T_{A_0}\mathcal{A}=\Omega^{1}(M,\mathfrak{g}).
\eeq{} 
The tangent spaces at points in the subspace of $\mathcal{M}$ determined by the other fields are also given by a space of differential forms, with the difference that for $\bar{a}_0$ and $b_0$ it is the space of harmonic zero-forms, whereas for $\alpha,c,\bar{c}$ it is the space of zero-forms. The determinant $h(\phi)$ is therefore given by
\beq
h(\phi)=\sqrt{\frac{\det_{\Omega^{0}(M,\mathfrak{g})}{L_{\phi}}}{\det_{\Omega^{1}(M,\mathfrak{g})}{L_{\phi}}}\cdot\frac{\det_{\Omega^{0}(M,\mathfrak{g})}{L_{\phi}}}{\det_{H^{0}(M,\mathfrak{g})}{L_{\phi}}}\cdot\frac{\det_{\Omega^{0}(M,\mathfrak{g})}{L_{\phi}}}{\det_{H^{0}(M,\mathfrak{g})}{L_{\phi}}}},
\eeq{}
where $H^{0}(M,\mathfrak{g})$ denotes the space of harmonic zero-forms with values in $\mathfrak{g}$. Since a one-form in three dimensions carries almost the same amount of degrees of freedom as three zero-forms, there is a huge cancelation between the above determinants; the final result will basically depend on the difference of zero-modes of differential forms of different degrees. To start with, the space of one-forms can be decomposed using the contact structure $\kappa$ into forms along $\kappa$ and the horizontal ones:
\beq
\Omega^{1}(M,\mathfrak{g})=\Omega^{1}_{\kappa}(M,\mathfrak{g})\oplus\Omega^{1}_{H}(M,\mathfrak{g}).
\eeq{}
A one-form $\tau$ can therefore be decomposed as
\beq
\tau=\kappa \tau_0 +\tau_{H}, \quad i_v \tau_H=0,
\eeq{}
where $\tau_0$ is a zero-form. Since the operator $L_{\phi}$ respects this decomposition, the determinant on the space of one-forms can be written
\beq
\text{det}_{\Omega^{1}(M,\mathfrak{g})}{L_{\phi}}=\text{det}_{\Omega^{0}(M,\mathfrak{g})}{L_{\phi}}\cdot \text{det}_{\Omega^{1}_{H}(M,\mathfrak{g})}{L_{\phi}},
\eeq{}
where the first factor is the determinant of the action on $\tau_0\in\Omega^{0}(M,\mathfrak{g})$ and the second one for $\tau_H\in\Omega^{1}_{H}(M,\mathfrak{g})$. Writing the determinant on the space of one-forms in this way, we can cancel out one factor in the numerator:
\beq
h(\phi)=\sqrt{\frac{1}{\det_{\Omega_{H}^{1}(M,\mathfrak{g})}{L_{\phi}}}\cdot\frac{\det_{\Omega^{0}(M,\mathfrak{g})}{L_{\phi}}}{\det_{H^{0}(M,\mathfrak{g})}{L_{\phi}}}\cdot\frac{\det_{\Omega^{0}(M,\mathfrak{g})}{L_{\phi}}}{\det_{H^{0}(M,\mathfrak{g})}{L_{\phi}}}}.
\eeq{}
Following the discussion in \cite{Beasley:2005vf}, we can Fourier expand horizontal forms into eigenmodes of the operator $\mathcal{L}_v$:
\beq
\xi=\sum_{t=-\infty}^{\infty}{\xi_{t}}, \quad \quad \xi\in\Omega_{H}^{\bullet}(M,\mathfrak{g}), \quad \quad \mathcal{L}_v\xi_t=2\pi i t \cdot\xi_t.
\eeq{}
The geometrical interpretation of $\xi_t$ is the following. If $\mathcal{L}$ denotes the line bundle over $\Sigma$ which describes the Seifert manifold $M$, $\xi_t$ can be thought of as sections of powers of $t$ of $\mathcal{L}$. Using the Fourier expansion, the spaces of zero-forms and horizontal one-forms can be decomposed into eigenspaces of $\mathcal{L}_v$ in the following way:
\beq
\begin{split}
\Omega^{1}_H(M,\mathfrak{g})&=\bigoplus_{t\neq 0}\Omega^{1}(\Sigma,\mathcal{L}^t\otimes \mathfrak{g})\bigoplus\Omega^{1}(\Sigma,\mathfrak{g}) \\
\Omega^{0}(M,\mathfrak{g})&=\bigoplus_{t\neq 0}\Omega^{0}(\Sigma,\mathcal{L}^t\otimes \mathfrak{g})\bigoplus\Omega^{0}(\Sigma,\mathfrak{g}), 
\end{split}
\eeq{}
where we have written the space which carries the trivial representation by itself. Using this decomposition, we find the determinant to be 
\beq
\left(\prod_{t\neq0}{\det{\left[\left(2\pi i t+[\phi,~]\right)_{|\mathfrak{g}}\right]^{\chi(\mathcal{L}^t)}}}\right)\cdot \det{\left([\phi,~]\right)_{|\mathfrak{g}}^{\frac{\chi(\Sigma)}{2}-\text{dim}H^{0}(M)}}.
\eeq{det1}
Here, the Euler character $\chi(\mathcal{L}^t)$ is due to the different amount of zero modes in the different spaces, and the remaning determinant is over the Lie algebra $\mathfrak{g}$. The last factor is one, since $\Sigma$ has genus zero. This factor is due to the the determinant arising from the constant ghosts $\bar{a}_0,b_0$ and the determinant coming from the space of trivial $U(1)$ representations of the other fields. 
\\
\\
We will now compute the determinant for the action on $\mathfrak{g}$. All fields transform in the adjoint representation of the gauge group, and $\phi$ lies in the Cartan subalgebra of $\mathfrak{g}$. We can decompose $\mathfrak{g}$ into root spaces:
\beq
\mathfrak{g}=\bigoplus_{\beta} \mathfrak{g}_{\beta},
\eeq{}
where $\beta$ denote the roots (we include the roots that are zero). The eigenvalues of $[\phi,~]$ acting on $\mathfrak{g}_{\beta}$ are given by $i\langle\beta,\phi\rangle$, where $\langle~,~\rangle$ denotes the pairing between $\mathfrak{g}$ and $\mathfrak{g}^*$. We therefore find the determinant to be 
\beq
\det\left(2\pi i t +[\phi,~]\right)=\prod_{\beta}\left(2\pi i t+i\langle\beta,\phi\rangle\right)=(2\pi i t)^{\Delta_{G}}\prod_{\beta>0}{\left(1-\left(\frac{\langle\beta,\phi\rangle}{2\pi t}\right)^2\right)}.
\eeq{det2}
In the above formula, $\Delta_{G}$ is the dimension of $G$, and we have used the fact that each non-zero root $\beta$ comes in a pair together with $-\beta$. Using $\eqref{det1}$ and $\eqref{det2}$, we find
\beq
h(\phi)=e^{-\frac{i\pi}{2}\eta}\prod_{t>0}{\left|\left(2\pi t\right)^{\Delta_{G}}\prod_{\beta>0}{\left(1-\left(\frac{\langle\beta,\phi\rangle}{2\pi t}\right)^2\right)}\right|^{\chi(\mathcal{L}^t)+\chi(\mathcal{L}^{-t})}}.
\eeq{hfinal}
At this point, our calculation has merged completely with the one in \cite{Beasley:2005vf}. The above expression is the one found in equation (5.50) there. Since the steps taken from now on will be identical to the ones in \cite{Beasley:2005vf}, we will simply quote their result below. For completeness, we review the main points in the calculation in appendix \ref{Details}. In that appendix we will also compute the phase of $h(\phi)$. Again, the original derivation of the phase is found in section 5 in \cite{Beasley:2005vf}, together with appendix C in \cite{Beasley:2009mb}. The phase factor can be computed since we know the eigenvalues of the operator $L_{\phi}$, and it generates the famous shift $k\rightarrow k+\check{c}_{\mathfrak{g}}$ of the Chern-Simons level $k$. $\check{c}_{\mathfrak{g}}$ is the dual Coxeter number of $\mathfrak{g}$.
\\
\\
The result we eventually obtain in appendix \ref{Details} is
\beq
\begin{split}
h(\phi)&=\exp{\left(-\frac{i\pi}{2}\eta_{0}(0)\right)}\times \exp{\left[\frac{i\check{c}_{\mathfrak{g}}}{4 \pi}\left(\frac{d}{P}\right)\text{Tr}\left(\phi\right)^{2}\right]} \\
&\times \frac{1}{P^{\Delta_{T}/2}}\prod_{\beta>0}{\langle\beta,\phi\rangle^{-2}\left[2\sin\left(\frac{\langle\beta,\phi\rangle}{2}\right) \right]^{2-N}\prod_{j=1}^{N}{\left[2\sin\left(\frac{\langle\beta,\phi\rangle}{2a_j}\right)\right]}}.
\end{split}
\eeq{}
Here, $\Delta_{T}$ is the dimension of the maximal torus $T$ of $G$. The first two factors arise from the computation of the phase. $\eta_{0}(0)$ is the part which is independent of $\phi$, given by
\beq
\eta_{0}(0)=-\frac{\Delta_{G}}{6}\cdot \frac{d}{P}+2\Delta_G\sum_{j=1}^{N}{s(b_j  ,a_j )}.
\eeq{}
The sum appearing above is the Dedekind sum, defined by
\beq
s(b,a)=\frac{1}{4a}\sum_{l=1}^{a-1}{\cot{\left(\frac{\pi l}{a}\right)}\cot{\left(\frac{\pi lb}{a}\right)}}.
\eeq{}
Plugging in this expression for $h(\phi)$ in $\eqref{Zres2}$, we find the final answer
\beq
\begin{split}
Z_{\{0\}}&=\exp{\left(-\frac{i\pi}{2}\eta_{0}(0)\right)}\cdot \frac{1}{P^{\Delta_{T}/2}}\frac{1}{\left|W\right |}\frac{1}{\text{Vol}(T)}\times \\
&\times \int_{\mathbf{t}}{[d\phi]~\exp{\left[\frac{i}{2\epsilon_{r}}\left(\frac{d}{P}\right)\text{Tr}\left(\phi^{2}\right)\right]}\prod_{\beta>0}{\left[2\sin\left(\frac{\langle\beta,\phi\rangle}{2}\right) \right]^{2-N}\prod_{j=1}^{N}{\left[2\sin\left(\frac{\langle\beta,\phi\rangle}{2a_j}\right)\right]}}}.
\end{split}
\eeq{Z0final} 
Here, we have introduced the renormalized coupling constant $\epsilon_r$:
\beq
\epsilon_r=\frac{2\pi}{k+\check{c}_{\mathfrak{g}}}.
\eeq{}
This is the same expression for the contribution of the trivial connection to the partition function on a Seifert homology sphere that was originally found in \cite{Marino:2002fk}, as a generalization of the results obtained in \cite{Lawrence:1999lr}. There it was found from working backwards of a complicated formula derived using the relation between Chern-Simons theory and two dimensional conformal field theory. It was later derived in a path integral computation using non-abelian localization in \cite{Beasley:2005vf}, and here we have derived it starting from the $\mathcal{N}=2$ supersymmetric version of Chern-Simons theory.  
\section{Wilson Loops} \label{Wilson section} 
As mentioned in section \ref{Localization}, we can include operators $\mathcal{O}$ in the path integral, and compute their expectation value exactly, as long as $Q\mathcal{O}=0$, where $Q$ is the odd operator defined in $\eqref{Q}$. A natural set of operators in Chern-Simons theory are Wilson loops, which are specified by a curve in $M$ and a representation of the gauge group $G$. The following Wilson loop along the fiber direction of the Seifert manifold fulfills the requirement of being $Q$-closed:
\beq
W_R=\text{Tr}_{R}\mathcal{P}\exp{\left(-\oint{\kappa\left(i_vA-i\sigma\right)}\right)}=\text{Tr}_{R}\mathcal{P}\exp{\left(\oint{\kappa\Phi}\right)},
\eeq{Wilson}
as one can easily check using the definition $\eqref{Q}$ of $Q$. $R$ denotes some representation of $G$, $\mathcal{P}$ denotes path ordering and we have written the operator both in terms of the original fields $A,\sigma$ and after we have made the change of variables $\sigma\rightarrow \Phi$, see equation $\eqref{cov}$. From the first expression, we see that this is a generalization of the Wilson loops considered in \cite{Kapustin:2009kz}. 
\\
\\
We now want to calculate the expectation value of this Wilson loop, that is computing the path integral
\beq
Z(k;R)=\frac{1}{\text{vol}(\mathcal{G})}\int{\mathcal{D}X~W_{R}~e^{iS}}.
\eeq{wilsonaux}
For clarity, the path integral is written before the gauge fixing, $X$ denotes all physical fields and $S$ is the action in $\eqref{twistaction}$. The calculation of $Z(k;R)$ is straightforward once we have computed the partition function. For example, if we consider the analogous situation as in section \ref{Calc}, namely the case of a Seifert homology sphere and localization at the trivial connection, we find without effort that the expectation value of the above defined Wilson loop in the fiber direction is given by
\beq
\begin{split}
&Z(k;R)_{\{0\}}=\exp{\left(-\frac{i\pi}{2}\eta_{0}(0)\right)}\cdot \frac{1}{P^{\Delta_{T}/2}}\frac{1}{\left|W\right |}\frac{1}{\text{Vol}(T)}\times\\
&\times \int_{\mathbf{t}}{[d\phi]\text{Tr}_{R}\exp{\left(\phi\right)}~\exp{\left[\frac{i}{2\epsilon_{r}}\left(\frac{d}{P}\right)\text{Tr}\left(\phi^{2}\right)\right]}\prod_{\beta>0}{\left[2\sin\left(\frac{\langle\beta,\phi\rangle}{2}\right) \right]^{2-N}\prod_{j=1}^{N}{\left[2\sin\left(\frac{\langle\beta,\phi\rangle}{2a_j}\right)\right]}}},
\end{split}
\eeq{Knot}
where we have reused the result $\eqref{Z0final}$, evaluated $\eqref{Wilson}$ on the solution $\eqref{trivialfp}$, used that $\oint{\kappa}=1$.
\\
\\
Of course, we know from \cite{Witten:1988hf} that expectation values of Wilson loops in Chern-Simons theory of the form
\beq
Z(k;R)=\frac{1}{\text{vol}(\mathcal{G)}}\int{\mathcal{D}A~\text{Tr}_{R}\mathcal{P}\exp{\left(-\oint{\kappa i_vA}\right)}~e^{ikS_{CS}}} 
\eeq{wilsonpure}
give knot invariants. As seen from the action in $\eqref{twistaction}$, the field $\sigma$ can be integrated out even in the presence of the Wilson loop $\eqref{Wilson}$ by performing the integral over the other auxiliary scalar $D$. Therefore the integral in $\eqref{wilsonaux}$ gives the same result as the integral $\eqref{wilsonpure}$. Moreover, perhaps somewhat surprisingly, as described in section 7 in \cite{Beasley:2009mb}, Wilson loops in the fiber direction of a Seifert manifold can actually give interesting knot invariants. For example, $S^{3}$ has many different descriptions as a Seifert fibration, not only the most well known one, that of a Hopf fibration. In the latter case, an expectation value of a Wilson loop in the fiber direction gives the unknot on $S^{3}$. This is the case considered in \cite{Kapustin:2009kz}. If we use a different description of $S^{3}$ as a Seifert manifold, also torus knots can be captured by these Wilson loops, see section 7 in \cite{Beasley:2009mb} for details. 
\\
\\
In the approach taken here, we can consider the same type of Wilson loops as in \cite{Beasley:2009mb}, and our result $\eqref{Knot}$ in the case of a Seifert homology sphere agrees exactly with the one computed in \cite{Beasley:2009mb}, equation (7.162) there.
\section{Chern-Simons theory on Seifert manifolds as a Cohomological TFT} \label{cohTFT}
An interesting aspect of the approach taken to Chern-Simons theory in this paper is the following. We know that the theory eventually localize to the space of solutions of the equations $\eqref{F}$, namely flat connections. There are in general two types of flat connections, reducible and irreducible. A connection is called reducible if there are non-trivial gauge transformations that fixes the connection. Otherwise the connection is called irreducible. For the case of irreducible flat connections, the only solution to the equations involving the auxiliary fields is that they vanish. Hence, the partition function, for a smooth component $\mathcal{F}$ of the moduli space of irreducible flat connections, can be written as an integral of an equivariant Euler class of the normal bundle to $\mathcal{F}$. Normally, Chern-Simons theory is considered as a topological field theory (TFT) of Schwarz type, as opposed to TFT's of Witten or Cohomological type. Cohomological type TFT's are constructed by specifying a set of fields, a set of equations and a set of symmetries, and the correlation functions of the theory constructed from this data compute intersection numbers on the moduli space of solutions to the equations modulo the symmetries. See for example \cite{Birmingham:1991ty} for a review. Since we have a set of equations to which the theory localizes, one would therefore guess that Chern-Simons theory on Seifert manifolds can be constructed in the same way as in this traditional approach to Cohomological TFT's. 
\\
\\
Actually, in \cite{Baulieu:1997nj} this approach to Chern-Simons theory on $\Sigma\times S^1$,  where $\Sigma$ is a Riemann surface, is taken. In that paper, a whole class of theories in different dimensions are constructed by specifying a set of fields, a set of equations and a set of symmetries. Then a transformation $Q$ is constructed in the standard way, used to impose the equations in the theory. More precisely, they start with a Cohomological TFT on some $D$ dimensional manifold, and then they attach a circle to obtain a $D+1$ dimensional theory. The symmetry in the $D$ dimensional theory is gauge symmetry, whereas in the $D+1$ dimensional theory the symmetry in question becomes gauge symmetry together with a $U(1)$ symmetry which corresponds to rotations along the circle. In three dimensions, they end up with a theory defined on the manifold $\Sigma\times S^1$, where $\Sigma$ is a Riemann surface. The monopole equations (corresponding to the first two equations in $\eqref{F}$) are chosen as the equations. Then some additional fields together with a transformation $Q$ are introduced in the canonical way to impose these equations in the theory. It is then shown that the Chern-Simons action (together with some additional fields) is a good \textit{observable}, good in the sense that it is $Q$-closed. It is then mentioned (but not shown in detail) that the correlation function of this observable actually is identical to the partition function of pure Chern-Simons theory on $\Sigma\times S^{1}$; that is, all fields except the gauge field can be considered as auxiliary in this case. It is also mentioned that the field content in this special case is the same as that of (partially twisted) $\mathcal{N}=1$ supersymmetric Yang-Mills in four dimensions, and the evaluation of the correlation function with the Chern-Simons action as an observable yields the known formula for the partition function of Chern-Simons theory on a manifold of the type $\Sigma\times S^1$. 
\\
\\
The situation we arrive at for Chern-Simons theory on Seifert manifolds is precisely analogous to the one in \cite{Baulieu:1997nj}. The field content is the same, since we originally started with the four dimensional $\mathcal{N}=1$ gauge multiplet. As already mentioned, the equations we impose, $\eqref{F}$, are the same. With the change of variable $\tilde{D}=-\frac{\kappa\wedge\ F}{\kappa\wedge d\kappa}+\frac{i}{2}\left(\sigma+D\right)$, the transformations $\eqref{auxsym}$ above are exactly the ones in \cite{Baulieu:1997nj}, given by equation (3.11) there. The vector field $v$ here corresponds to $\frac{\partial}{\partial t}$ there, where $t$ is the coordinate along $S^{1}$. Moreover, the action we consider in this paper, given in $\eqref{twistaction}$, is (with the change of variable $D\rightarrow \tilde{D}$) the same as the one which is shown to be a good observable in \cite{Baulieu:1997nj}, given by equation (4.4) there. (The extra terms in our action compared to the one in \cite{Baulieu:1997nj} is due to the fact that $dt$ is not a contact structure, but it corresponds to $\kappa$ here. Since $d\kappa\neq 0$, as opposed to $d(dt)=0$, we need a few extra terms in our action to obtain $Q$ invariance.) Also the Wilson loops which we consider, equation $\eqref{Wilson}$, are the same as the ones in \cite{Baulieu:1997nj}, equation (4.2) there.
\\
\\
Given the above, it would be interesting to see how the construction of the Cohomological TFT's in \cite{Baulieu:1997nj} changes when non-trivial $S^{1}$ fibrations are considered.
\section{Summary and discussion}
In this article we have studied path integral localization of Chern-Simons theory. The main result is the introduction of the auxiliary fields and odd symmetries in $\eqref{auxsym}$. These transformations can be defined on any compact, orientable three manifold $M$ with a choice of contact structure. The operator generating the symmetry has a natural interpretation as an equivariant differential on the space of fields and, when $M$ is a Seifert manifold, we have shown that the symmetry can be used to reduce the calculation of the partition function to an integral over the moduli space of flat connections together with an integral over an auxiliary covariantly constant scalar field. This framework is a generalization of the one introduced by Kapustin, Willett and Yaakov in \cite{Kapustin:2009kz}, and it is obtained via a variant of topological twisting. We have also discussed the incorporation of Wilson loops along the fiber direction of the Seifert manifold, and shown how to compute their expectation values exactly within this framework. 
\\
\\
As for explicit calculations, we have focused our attention to the contribution to the partition function and expectation values of Wilson loops arising from a point in the space of flat connections corresponding to an isolated, trivial connection. We have found complete agreement with the results obtained by other methods, for example the method of non-abelian localization in \cite{Beasley:2005vf,Beasley:2009mb}. The framework presented here is not restricted to this case, and can be applied in more general situations. In the general case, the gauge fixing in section \ref{gaugefix} should be done with a non-trivial background connection. We must also know the value of the Chern-Simons action for a non-trivial flat connection, which for many cases can be found in \cite{Rozansky:1993zx} or from references therein. 
\\
\\
The main difference between this work and \cite{Kapustin:2009kz}, apart from the topological twisting, is the following. The supersymmetry used in \cite{Kapustin:2009kz} squares to zero, whereas the symmetry we use here squares to a symmetry of the theory. This aspect simplifies the calculation of the one-loop determinants appearing in the localization calculation. For example, the phase of the determinant was not computed in \cite{Kapustin:2009kz}, whereas here it can be computed.
\\
\\
The perhaps most interesting aspect of this approach to Chern-Simons theory is the relation to the standard approach to Cohomological TFT's. As we have shown in section \ref{cohTFT}, there is a clear relation between the approach taken here and the one taken by Baulieu, Losev and Nekrasov in \cite{Baulieu:1997nj}, where Chern-Simons theory on $\Sigma\times S^{1}$ is handled as a Cohomological TFT. It would be interesting to investigate further Chern-Simons theory on Seifert manifolds from this perspective. 
\\
\\
\bigskip
\noindent{\bf\large Acknowledgement}: 
\\
The author would like to thank M. Zabzine for suggesting the project and for many interesting discussions and useful comments. 
\bigskip
%\pagebreak
\appendix
\section{The topological twist} \label{twist}
In this appendix, we will derive the action $\eqref{twistaction}$ and transformations $\eqref{auxsym}$ starting from the supersymmetry transformations $\eqref{s3action}$ and the supersymmetric action  $\eqref{susy}$. The easiest way to do this is to introduce a metric on the manifold $M$ on which the theory is defined, even though the final result holds on any compact, orientable three manifold without a choice of metric, as can be easily checked. In the derivation, we will need to use a few structures which is present on contact manifolds, which we now describe.
\subsection{More basic facts about contact manifolds}
In order to twist the theory, we have to temporarily choose a metric on $M$. With the metric, we can define a Hodge star operator $*$. On a three dimensional contact manifold we can always choose a metric that fulfills
\beq
*1=\kappa\wedge d\kappa, \quad \quad *\kappa=d\kappa.
\eeq{}
Given the contact structure, we can separate differential forms on $M$ into those along $\kappa$, and those which are horizontal:
\beq
\Omega^{\bullet}=\Omega^{\bullet}_\kappa\oplus\Omega^{\bullet}_{H}.
\eeq{}
Moreover, let us define an operator $J$ which will be important in the topological twist performed in the next subsection. It acts on differential forms and is given by:
\beq
J=-i_v \;*.
\eeq{cs}
When acting on horizontal one-forms, $J$ fulfills $J^{2}=-1$ and therefore defines a complex structure on $\Omega^{1}_H$. Using the complex structure $J$, we can define projectors 
\beq
P_{\pm}=\frac{1}{2}\left(1\mp i\;J\right).
\eeq{projector}
We can now further decompose $\Omega^{1}_{H}$ into $\pm i$ eigenspaces of $P_{\pm}$:
\beq
\Omega^{1}_H=\Omega^{(1,0)}_H\oplus\Omega^{(0,1)}_H.
\eeq{} 
\subsection{The twist}
The spinor $\epsilon$ used when localizing on $S^{3}$, equation $\eqref{killingspinor}$, defines a vector field $v^{\mu}$ on $M$ through the relation
\beq
\epsilon^{\dagger}\gamma^{\mu}\epsilon=v^{\mu}.
\eeq{reebs3}
It is noted in \cite{Kapustin:2009kz} that this vector field generates the fiber of the Hopf fibration of $S^3$. This vector field is therefore the Reeb vector field in the case of $S^{3}$. We will use this observation in order to define auxiliary fields and odd transformations using differential forms instead of spinors.
\\
\\
It turns out that for our considerations, it is better to choose a slightly different supersymmetry as a starting point, as compared to the one used in \cite{Kapustin:2009kz}. We use the transformations $\eqref{susy}$ with spinors defined by 
\beq
\begin{split}
\nabla_{\mu}\epsilon&=\frac{i}{2}\gamma_{\mu}\epsilon, \quad \quad \quad \epsilon^{\dagger}\epsilon=1 \\
\eta&=\epsilon.
\end{split}
\eeq{oursusy}
In order to perform the twist, consider an odd, Lie algebra valued one-form $\rho\in\Omega^{1}_\kappa(M,\mathfrak{g})\oplus\Omega^{(1,0)}_H(M,\mathfrak{g})$ together with its complex conjugate $\bar{\rho}\in\Omega^{1}_\kappa(M,\mathfrak{g})\oplus\Omega^{(0,1)}_H(M,\mathfrak{g})$. We can decompose these one-forms as
\beq
\rho=\kappa \rho_0+\rho_{(1,0)}, \quad \quad \bar{\rho}=\kappa\bar{\rho_0}+\rho_{(0,1)}.
\eeq{} 
The odd one-forms $\rho_{(1,0)}$ and $\rho_{(0,1)}$ by definition fulfill the following relations
\beq
\begin{split}
i_v\rho_{(1,0)}&=i_v\rho_{(0,1)}=0, \\
P_{+}\rho_{(1,0)}=\rho_{(1,0)}, \quad P_-\rho_{(1,0)}&=0, \quad P_{+}\rho_{(0,1)}=0, \quad P_-\rho_{(0,1)}=\rho_{(0,1)},
\end{split}
\eeq{chihor}
where $v$ is the vector field dual to $\kappa$, and $P_\pm$ are the projectors defined in $\eqref{projector}$.
\\
\\
Let now $\rho$ and $\bar{\rho}$ be defined in terms of the spinors $\lambda$ and $\lambda^{\dagger}$ appearing in the $\mathcal{N}=2$ supersymmetric Chern-Simons theory:
\beq
\lambda=\gamma^{\mu}\epsilon\rho_{\mu}, \quad \quad \lambda^{\dagger}=\epsilon^{\dagger}\gamma^{\mu}\bar{\rho}_\mu.
\eeq{mapspinorform}
$\epsilon$ is the spinor fulfilling the relations $\eqref{oursusy}$. We now use the fact that the Pauli matrices $\gamma_{\mu}$ fulfill the relation
\beq
\gamma_{\mu}\gamma_{\nu}=\delta_{\mu\nu}+i\epsilon_{\mu\nu\rho}\gamma^{\rho},
\eeq{paulirel}
and that the contact structure fulfills
\beq
i_v\kappa=1, \quad \quad i_v d\kappa=0.
\eeq{}
These two relations, together with equations $\eqref{oursusy},\eqref{cs},\eqref{reebs3},\eqref{chihor}$ allows us to solve for $\rho_{0},\bar{\rho_{0}},\rho_{(1,0)},\rho_{(0,1)}$ in $\eqref{mapspinorform}$: 
\beq
\begin{split}
\rho_0=\epsilon^{\dagger}\lambda, \quad \quad \bar{\rho_{0}}=\lambda^{\dagger}\epsilon, \quad \quad \rho_{(1,0)}=-\frac{i}{2}J\epsilon^{\dagger}\gamma\lambda, \quad \quad \rho_{(0,1)}=\frac{i}{2}J\lambda^{\dagger}\gamma\epsilon.
\end{split}
\eeq{}
Written in the new variables $\rho_{0},\bar{\rho_{0}},\rho_{(1,0)},\rho_{(0,1)}$, the transformations $\eqref{susy}$ for the supersymmetric Chern-Simons theory become
 \beq
\begin{split}
\delta A&=\frac{i}{2}\left(\kappa\bar{\rho_{0}}+2\rho_{(0,1)}\right) + \frac{i}{2}\left(\kappa\rho_{0}+2\rho_{(1,0)}\right)\\
\delta \sigma&=\frac{1}{2}\bar{\rho_{0}}+\frac{1}{2}\rho_{0} \\
\delta D&=i\mathcal{L}_v^{A}\bar{\rho_{0}}+2\frac{1}{\kappa\wedge d\kappa}\left(\kappa\wedge d_{A}\rho_{(0,1)}\right)+\frac{1}{2}\bar{\rho_{0}}-[\bar{\rho_{0}},\sigma] \\
& -i\mathcal{L}_v^{A}\rho_{0}+2\frac{1}{\kappa\wedge d\kappa}\left(\kappa\wedge d_{A}\rho_{(1,0)}\right)+\frac{1}{2}\rho_{0}+[\rho_{0},\sigma] \\
\delta \rho_{0}&=i\frac{1}{\kappa\wedge d\kappa}\left(\kappa\wedge F\right)+\frac{1}{2}\left(\sigma+D\right)-i~\mathcal{L}^{A}_v \sigma \\
\delta\bar{\rho_{0}}&=-i\frac{1}{\kappa\wedge d\kappa}\left(\kappa\wedge F\right)-\frac{1}{2}\left(\sigma+D\right)-i~\mathcal{L}^{A}_v \sigma \\
\delta \rho_{(1,0)}&=-P_{+}\left (i_v F+J\;d_{A}\sigma\right) \\
\delta\rho_{(0,1)}&=-P_{-}\left (i_v F-J\;d_{A}\sigma\right).  
\end{split}
\eeq{bigaux}
Let us now introduce the odd one-form $\Psi\in\Omega^{1}(M,\mathfrak{g})$ and the odd zero-form $\alpha\in\Omega^{0}(M,\mathfrak{g})$, defined as
\beq
\begin{split}
\Psi&=\frac{i}{2}\kappa\left(\rho_{0}+\bar{\rho_{0}}\right)+i\left(\rho_{(1,0)}+\rho_{(0,1)}\right) \\
\alpha&=\frac{i}{2}\left(\rho_{0}-\bar{\rho_{0}}\right).
\end{split}
\eeq{}
In terms of $\Psi$ and $\alpha$, the transformations $\eqref{bigaux}$ reduce to\footnote{In the final form of the auxiliary transformations presented below we have also rearranged a few factors of ÓiÓ, as compared to $\eqref{bigaux}$.}
\beq
\begin{split}
\delta A&=\Psi \\
\delta \Psi&=i_v F+id_{A}\sigma \\
\delta \alpha &= -\frac{\kappa\wedge F}{\kappa\wedge d\kappa}+\frac{i}{2}\left(\sigma+D\right)\\
\delta\sigma&=-i~i_v\Psi \\
\delta D&=-2i\mathcal{L}^{A}_v\alpha-2[\sigma,\alpha]-2i\frac{\kappa\wedge d_{A}\Psi}{\kappa\wedge d\kappa}+i~i_v\Psi
\end{split}
\eeq{auxsym2}
and the action $\eqref{s3action}$ becomes
\beq
\begin{split}
S&=\frac{k}{4\pi}\int_M{\text{Tr}\left(A\wedge dA+\frac{2}{3}A\wedge A\wedge A-\kappa\wedge \Psi\wedge \Psi-2d\kappa\wedge\Psi\alpha+\kappa\wedge d\kappa~D\sigma\right)}. \\
\end{split}
\eeq{}
It is straightforward to verify that the transformation $\delta$ in $\eqref{auxsym2}$ squares to a translation along the Reeb vector field plus a gauge transformation with parameter $\Phi=i\sigma-i_v A$ using only properties which any contact structure has.
\section{Details of computation of the determinant} \label{Details}
In this appendix we will show the main steps in simplifying the expression in equation $\eqref{hfinal}$,  which we reproduce here:
\beq
h(\phi)=e^{-\frac{i\pi}{2}\eta}\prod_{t>0}{\left|\left(2\pi t\right)^{\Delta_{G}}\prod_{\beta>0}{\left(1-\left(\frac{\langle\beta,\phi\rangle}{2\pi t}\right)^2\right)}\right|^{\chi(\mathcal{L}^t)+\chi(\mathcal{L}^{-t})}}.
\eeq{happ}
The same calculation is performed in \cite{Beasley:2005vf}, which we will closely follow. The Riemann zeta-function $\zeta(s)$ will play an important role in the computations below. It is defined by
\beq
\zeta(s)=\sum_{t>0}{\frac{1}{t^s}}.
\eeq{}
Here $s$ is a complex parameter. When the real part of $s$ is greater than one the sum converges and for other values of $s$ $\zeta(s)$ is defined by analytic continuation. The properties of $\zeta(s)$ that we will need are
\beq
\begin{split}
\zeta(0)&=-\frac{1}{2} \\
\zeta(-1)&=-\frac{1}{12} \\
\zeta'(0)&=-\frac{1}{2}\ln(2\pi)
\end{split}
\eeq{zeta0}
and that it behaves as
\beq
\zeta(s+1)=\frac{1}{s}+\gamma_{0}+s\gamma_{1}\ldots
\eeq{}
as $s$ approaches zero. We will also need the Hurwitz zeta function $\zeta(s,a)$, defined by
\beq
\zeta(s,a)=\sum_{m=0}^{\infty}{\frac{1}{(m+a)^{s}}}.
\eeq{}
Here $a$ is a real parameter in the interval $0<a\leq 1$, and $s$ is as above a complex variable. The property of $\zeta(s,a)$ that we will need is
\beq
\zeta(0,a)=\frac{1}{2}-a.
\eeq{}
\subsection{Calculation of the absolute value}
The first thing to do is to simplify the exponent in $\eqref{happ}$. If $\text{deg}(\mathcal{L})=n$, then the Riemann-Roch theorem tells us that
\beq
\chi(\mathcal{L})=n+1-g,
\eeq{}
where $g$ is the genus of the Riemann surface. Similarly, for $\mathcal{L}^t$ we have
\beq
\chi(\mathcal{L}^t)=\text{deg}(\mathcal{L}^t)+1-g.
\eeq{riemannroch}
Using the above relation and remembering that $g=0$ in our case we find
\beq
\chi(\mathcal{L}^t)+\chi(\mathcal{L}^{-t})=\text{deg}(\mathcal{L}^t)+\text{deg}(\mathcal{L}^{-t})+2.
\eeq{}
Since we are working on an orbifold line bundle, the degree is not multiplicative, so the above expression is not just $2$. Instead, it is shown in a few steps in \cite{Beasley:2005vf} that
\beq
\chi(\mathcal{L}^t)+\chi(\mathcal{L}^{-t})=2-N+\sum_{j=1}^{N}{\varphi_{a_j }(t)}.
\eeq{chi+chi}
Here $\varphi_{a_j }(t)$ is a function which takes the value 1 if $a_j$ divides $t$, and zero otherwise:
\beq
\varphi_{a_j }(t)= \begin{array}{ll}
					1 & \text{if   } \quad a_j|t \\
					0 & \text{otherwise}
				\end{array} .
\eeq{}
For any function $f(t)$, the following identity holds:
\beq
\prod_{t>0}{f(t)^{\varphi_{a_j }(t)}}=\prod_{t>0}{f(a_j\cdot t)}.
\eeq{prodphi}
Using $\eqref{chi+chi}$ and $\eqref{prodphi}$, we find
\beq{}
\begin{split}
h(\phi)&=e^{-\frac{i\pi}{2}\eta}\prod_{t>0}{\left|\left(2\pi t\right)^{\Delta_{G}}\prod_{\beta>0}{\left(1-\left(\frac{\langle\beta,\phi\rangle}{2\pi t}\right)^2\right)}\right|^{2-N+\sum_{j=1}^{N}{\varphi_{a_j }(t)}}} \\
&=e^{-\frac{i\pi}{2}\eta}f(\phi)^2\cdot \prod_{j=1}^{N}{\left|\frac{f_{a_j}(\phi)}{f(\phi)}\right|}
\end{split}
\eeq{h2}
where
\beq
\begin{split}
f(\phi)&=\prod_{t>0}{\left[(2\pi t)^{\Delta_{G}}\prod_{\beta>0}{\left(1-\left(\frac{\langle\beta,\phi\rangle}{2\pi t}\right)^{2}\right)}\right]} \\
f_{a_j}(\phi)&=\prod_{t>0}{\left[(2\pi t\cdot a_j)^{\Delta_{G}}\prod_{\beta>0}{\left(1-\left(\frac{\langle\beta,\phi\rangle}{2\pi t\cdot a_j}\right)^{2}\right)}\right]}.
\end{split}
\eeq{f}
We now use the identity
\beq
\frac{\sin(x)}{x}=\prod_{t>0}{\left(1-\left(\frac{x}{\pi t}\right)^2\right)}
\eeq{}
together with $\zeta(s)$ to evaluate the products over $t$ in $\eqref{f}$:
\beq
\begin{split}
f(\phi)&=\prod_{t>0}{(2\pi )^{\Delta_{G}}}\cdot \prod_{t>0}{t^{\Delta_{G}}}\cdot \prod_{t>0}{\left[\prod_{\beta>0}{\left(1-\left(\frac{\langle\beta,\phi\rangle}{2\pi t}\right)^{2}\right)}\right]} \\
&=(2\pi)^{-\Delta_{G}/2}\cdot (2\pi)^{\Delta_{G}/2}\cdot \prod_{\beta>0}{\frac{2}{\langle \beta,\phi\rangle}\sin\left(\frac{\langle\beta,\phi\rangle}{2}\right)} \\
&=\prod_{\beta>0}{\frac{2}{\langle\beta,\phi\rangle}\sin\left(\frac{\langle\beta,\phi\rangle}{2}\right)} \\
f_{a_j}(\phi)&=\prod_{t>0}{(2\pi \cdot a_j)^{\Delta_{G}}}\cdot \prod_{t>0}{t^{\Delta_{G}}}\cdot {\left[\prod_{\beta>0}{\left(1-\left(\frac{\langle\beta,\phi\rangle}{2\pi t\cdot a_j}\right)^{2}\right)}\right]} \\
&=(2\pi\cdot a_j)^{-\Delta_{G}/2}\cdot (2\pi)^{\Delta_{G}/2}\cdot \prod_{\beta>0}{\frac{2\cdot a_j}{\langle \beta,\phi\rangle}\sin\left(\frac{\langle\beta,\phi\rangle}{2\cdot a_j}\right)} \\
&=a_j^{-\Delta_{T}/2}\cdot \prod_{\beta>0}{\frac{2}{\langle\beta,\phi\rangle} \sin\left(\frac{\langle\beta,\phi\rangle}{2\cdot a_j}\right)} .
\end{split}
\eeq{f2}
Above, we have used that $\zeta(0)=-\frac{1}{2} \text{ and } \zeta'(0)=-\frac{1}{2}\ln(2\pi)$, and $\Delta_T=\text{dim T}$. Plugging $\eqref{f2}$ into $\eqref{h2}$, we find
\beq
h(\phi)=e^{-\frac{i\pi}{2}\eta}\cdot\frac{1}{P^{\Delta_{T}/2}}\cdot \prod_{\beta>0}{\langle\beta,\phi\rangle^{-2}\left|2\sin\left(\frac{\langle\beta,\phi\rangle}{2}\right)\right|^{2-N}\cdot\prod_{j=1}^{N}{\left|2\sin\left(\frac{\langle\beta,\phi\rangle}{2\cdot a_j}\right)\right|}},
\eeq{}
where $P=\prod_{j=1}^{N}{a_j}$.
\subsection{Calculation of the phase}
A regularized version of $\eta$ in $\eqref{eta}$ is given by
\beq
\begin{split}
\eta_{\phi}(s)&=\sum_{t=-\infty}^{\infty}\sum_{\beta}{\chi(\mathcal{L}^t)\text{sign}(\lambda(t,\beta))\left|\lambda(t,\beta)\right|^{-s}} \\
\lambda(t,\beta)&=2\pi t+\langle\beta,\phi\rangle.
\end{split}
\eeq{}
In the above formula, the multiplicities eigenvalues are taken into account by the Euler character $\chi(\mathcal{L}^t)$, and $\lambda(t,\beta)$ are the eigenvalues of our operator $i\mathcal{L}_v+i[\phi,~]$. The eigenvalue $\lambda(t,\beta)=0$ is not included in the sum. We want to compute $\eta\equiv\eta_{\phi}(0)$. It turns out to be convenient to split $\eta_{\phi}(s)$ in order to isolate the $\phi$-independent part:
\beq
\delta\eta_{\phi}(s)=\eta_{\phi}(s)-\eta_0(s),
\eeq{}
where $\eta_{0}(s)$ is defined by
\beq
\eta_{0}(s)=\sum_{t\neq 0}\sum_{\beta}{\chi(\mathcal{L}^t) \text{sign}(2\pi t)\left|2\pi t\right|^{-s}}.
\eeq{}
We start with this quantity. It can be rewritten as 
\beq
\eta_{0}(s)=\sum_{t>0}\sum_\beta{\frac{\chi(\mathcal{L}^{t})-\chi(\mathcal{L}^{-t})}{(2\pi t)^s}}.
\eeq{eta0}
Using equation $\eqref{riemannroch}$, we find
\beq
\chi(\mathcal{L}^{t})-\chi(\mathcal{L}^{-t})=\text{deg}(\mathcal{L}^{t})-\text{deg}(\mathcal{L}^{-t}).
\eeq{}
Again, to evaluate this quantity takes some work. All the details can be found in the appendix of \cite{Beasley:2009mb}, and the final answer is
\beq
\chi(\mathcal{L}^{t})-\chi(\mathcal{L}^{-t})=2t\cdot \frac{d}{P}-2\sum_{j=1}^{N}{\left(\left(\frac{tb_j  }{a_j}\right)\right)}
\eeq{chi-chi}
where 
\beq
\left(\left(x\right)\right)=\begin{array}{ll}
					\{x\}-\frac{1}{2} & \text{if } x\in\mathbb{R}Š-\mathbb{Z} \\
					0 & \text{if } x\in\mathbb{Z}
				\end{array}.
\eeq{x}
$\{x\}$ denotes the element in the interval $[0,1)$ such that $x-\{x\}\in\mathbb{Z}$. We notice that $\left(\left(x+1\right)\right)=\left(\left( x\right)\right)$ and hence we have
\beq
\left(\left(\frac{tb_j  }{a_j}\right)\right)=\left(\left(\frac{(t+ma_j)b_j  }{a_j}\right)\right), \quad m\in\mathbb{Z},
\eeq{shift}
which we will use later. Now, plugging in $\eqref{chi-chi}$ in $\eqref{eta0}$ we get 
\beq
\eta_0(s)=\frac{2\Delta_G}{(2\pi)^s}\cdot\frac{d}{P}\zeta(s-1)-\frac{2\Delta_G}{(2\pi)^s}\cdot \sum_{j=1}^{N}\sum_{t>0}{\frac{1}{t^s}\left(\left(\frac{tb_j  }{a_j}\right)\right)}.
\eeq{eta3}
Above, we have identified the Riemann zeta-function in the first term, and the factor $\Delta_G$ comes from the sum over $\beta$. The last term can be rewritten as
\beq
\sum_{t>0}{\frac{1}{ t^s}\left(\left(\frac{tb_j  }{a_j}\right)\right)}=\sum_{l=1}^{a_j}\sum_{m=0}^{\infty}{\frac{1}{ (l+ma_j)^s}\left(\left(\frac{(l+ma_j)b_j  }{a_j}\right)\right)}=\sum_{l=1}^{a_j}{\left(\left(\frac{lb_j  }{a_j}\right)\right)\frac{1}{a^s_j}\zeta(s,\frac{l}{a_j})}.
\eeq{}
In the last step we have used $\eqref{shift}$ and identified the Hurwitz zeta-function. Using this in $\eqref{eta3}$, plugging in the values $\zeta(-1)=-\frac{1}{12}$ and $\zeta(0,\frac{l}{a_j})=\frac{1}{2}-\frac{l}{a_j}=-\left(\left(\frac{l}{a_j}\right)\right)$, we find
\beq
\eta_0(0)=-\frac{\Delta_G}{6}\cdot\frac{d}{P}+2\Delta_G\cdot \sum_{j=1}^{N}\sum_{l=1}^{a_j}{\left(\left(\frac{lb_j  }{a_j}\right)\right)\left(\left(\frac{l}{a_j}\right)\right)}.
\eeq{}
One can show that the last term is nothing but the Dedekind sum, as defined in the main text, so we finally get
\beq
\eta_{0}(0)=-\frac{\Delta_{G}}{6}\cdot \frac{d}{P}+2\Delta_G\sum_{j=1}^{N}{s(b_j  ,a_j )}.
\eeq{eta02}
As for $\delta\eta_{\phi}(s)$, it can be written as 
\beq
\begin{split}
\delta\eta_{\phi}(s)&=\eta_{\phi}(s)-\eta_{0}(s) \\
&=\sum_{t>0}\sum_{\beta>0}{\left(\chi(\mathcal{L}^{t})-\chi(\mathcal{L}^{-t})\right)\left(\frac{1}{\left(2\pi t+\langle\beta,\phi\rangle\right)^s}-\frac{1}{(2\pi t)^s}\right)} \\
&+\sum_{t>0}\sum_{\beta>0}{\left(\chi(\mathcal{L}^{t})-\chi(\mathcal{L}^{-t})\right)\left(\frac{1}{\left(2\pi t-\langle\beta,\phi\rangle\right)^s}-\frac{1}{(2\pi t)^s}\right)} \\
&+\chi(\mathcal{L}^{0})\sum_{\beta\neq 0}{\text{sign}(\langle\beta,\phi\rangle)|\langle\beta,\phi\rangle|^{-s}}.
\end{split}
\eeq{}
Here we have without loss assumed that
\beq
0<\frac{\langle\beta,\phi\rangle}{2\pi}<1.
\eeq{}The last term is zero since for every root $\beta$ there is a corresponding root $-\beta$. If we expand the above expression around $s=0$, the first non-trivial term in the expansion will cancel between the two sums, the second term will add between the two sums, and higher order terms will vanish in the $s\rightarrow 0$ limit. Plugging in $\eqref{chi-chi}$ we find
\beq
\delta\eta_{\phi}(s)=2\cdot\frac{d}{P}\cdot\frac{1}{(2\pi)^{s}}\sum_{t>0}\sum_{\beta>0}{s(s+1)\left(\frac{\langle\beta,\phi\rangle}{2\pi}\right)^{2}\frac{1}{t^{s+1}}}+\sum_{t>0}\sum_{\beta>0}{s\cdot \mathcal{O}\left(\frac{1}{t^{s+2}}\right)}.
\eeq{}
To convince oneself that the second term in $\eqref{chi-chi}$ will give no contribution in the limit $s\rightarrow 0$, one can follow the same steps as in the calculation of $\eta_{0}(s)$. In the first term, we identify $\zeta(s+1)$, which nicely cancels out the factor of $s$ at $s=0$, and we get
\beq
\delta\eta_{\phi}(0)=2\cdot\frac{d}{P}\sum_{\beta>0}{\left(\frac{\langle\beta,\phi\rangle}{2\pi}\right)^{2}}=\frac{1}{2\pi^{2}}\cdot\frac{d}{P}\sum_{\beta>0}{\langle\beta,\phi\rangle}^2=-\frac{\check{c}_{\mathfrak{g}}}{2\pi^{2}}\cdot\frac{d}{P}\text{Tr}(\phi)^2
\eeq{}
where in the last step we have used the formula
\beq
\sum_{\beta>0}{\langle\beta,\phi\rangle}^2=-\check{c}_{\mathfrak{g}}\text{Tr}(\phi)^2
\eeq{deltaeta}
which holds for any simply-laced group. Combing $\eqref{eta0}$ and $\eqref{deltaeta}$ we find the final form of the phase of the determinant:
\beq
\eta=-\frac{\Delta_{G}}{6}\cdot \frac{d}{P}+2\Delta_G\sum_{j=1}^{N}{s(b_j  ,a_j )}-\frac{\check{c}_{\mathfrak{g}}}{2\pi^{2}}\cdot\frac{d}{P}\text{Tr}(\phi)^2.
\eeq{}
When combined with the absolute value of the determinant, the last term above gives rise to the quantum shift $k\rightarrow k+\check{c}_{\mathfrak{g}}$ of the Chern-Simons level.
%\appendixpage

\bibliographystyle{utphys}
\bibliography{Localization}

\end{document}